\documentclass[showpacs,preprintnumbers,amsmath,amssymb]{revtex4}
\usepackage{graphicx}% Include figure files
\usepackage{dcolumn}% Align table columns on decimal point
\usepackage{bm}% bold math

\begin{document}

\preprint{KUNS-2213}

\title{Constraining alternative theories of gravity by gravitational waves \\from precessing eccentric compact binaries with LISA}% Force line breaks with \\

\author{Kent Yagi}
%\author{Takahiro Tanaka}%
\affiliation{%
Department of Physics, Kyoto University,
   Kyoto, 606--8502, Japan%\\
}%

\author{Takahiro Tanaka}
\affiliation{
Yukawa Institute for Theoretical Physics,
  Kyoto University,
  Kyoto 606--8502, Japan%\\
}%

\date{\today}% It is always \today, today,

\begin{abstract}
\quad We calculate how strongly one can put constraints on alternative theories of gravity such as Brans-Dicke and massive graviton theories with LISA. 
We consider inspiral gravitational waves from a compact binary composed of a neutron star (NS) and an intermediate mass black hole (IMBH) in Brans-Dicke (BD) theory and that composed of 2 super massive black holes (SMBHs) in massive graviton theories. 
We use the restricted 2PN waveforms including the effects of spins. 
We also take both precession and eccentricity of the orbit into account.
For simplicity, we set the fiducial value for the spin of one of the binary constituents to zero so that we can apply the approximation called \textit{simple precession}. 
We perform the Monte Carlo simulations of $10^4$ binaries, estimating the determination accuracy of binary parameters including the BD parameter $\omega_{\mathrm{BD}}$ and the Compton wavelength of graviton $\lambda_g$ for each binary using the Fisher matrix method. 
We find that including both the spin-spin coupling $\sigma$ and the eccentricity $e$ into the binary parameters reduces the determination accuracy by an order of magnitude for the Brans-Dicke case, whilst it has less influence on massive graviton theories. 
On the other hand, including precession enhances the constraint on $\omega_{\mathrm{BD}}$ only 20$\%$ but it increases the constraint on $\lambda_g$ by an order of magnitude. 
Using a $(1.4+1000)M_{\odot}$ NS/BH binary of SNR=$\sqrt{200}$, one can put a constraint $\omega_{\mathrm{BD}}>6944$, whilst using a $(10^7+10^6)M_{\odot}$ BH/BH binary at 3Gpc, one can put $\lambda_g>3.06\times10^{21}$cm, on average. 
The latter is 4 orders of magnitude stronger than the one obtained from the solar system experiment. 
These results are consistent with previous results within uncontrolled errors and indicate that the effects of precession and  eccentricity must be taken carefully in the parameter estimation analysis.
\end{abstract}

\pacs{Valid PACS appear here}% PACS, the Physics and Astronomy
\maketitle

\section{INTRODUCTION}

\quad Recent observations (e.g. type Ia SNe~\cite{riess}) show that the expansion of the Universe is accelerating at the moment.
%However, no one knows the origin of this expansion.
One possible way to explain this accelerated expansion is to introduce some undiscovered matter field such as quintessence or k-essence (for a recent review, see~\cite{copeland}).
Another possibility is to modify gravitational theory from general relativity.
The easiest modification is to add some scalar degrees of freedom to gravity.
This type of theory is called the scalar-tensor theory~\cite{fujii}.
This theory also appears to be a candidate for solving the inflation problem, known as the hyper-extended inflation~\cite{steinhardt}.
Furthermore, scalar fields called dilatons and moduli may play the role of these scalar degrees of freedom in the context of the string theory.
Therefore it is very important to put a strong observational constraint on this theory.
In this paper, we focus on Brans-Dicke (BD) theory~\cite{brans} as a representative of scalar-tensor theory.
This theory has the so-called Brans-Dicke parameter $\omega_{\mathrm{BD}}$ and by taking the limit $\omega_{\mathrm{BD}}\rightarrow\infty$, it reduces to general relativity.
The current strongest constraint on the Brans-Dicke parameter is $\omega_{\mathrm{BD}}>40000$, which is obtained from the solar system experiment, measuring the Shapiro time delay using the Saturn probe satellite Cassini~\cite{cassini}.

Another type of modified theory of gravity introduces a finite mass $m_g$ to a graviton (see~\cite{rubakov} for a recent review).
This type of theory is called the massive graviton theory.
Originally, Fierz and Pauli~\cite{fierz} proposed a Lorentz-invariant massive gravity by simply adding a graviton mass term to the Einstein-Hilbert action at the quadratic level. 
However, it was found that the linearised Fierz-Pauli theory does not approach linearised general relativity in the massless limit. 
This is the so-called van Dam-Veltman-Zakharov (vDVZ) discontinuity~\cite{vdv,z}.
This is due to the fact that the helicity-0 component of the graviton does not decouple from matter.
This discontinuity seems to contradict with solar system experiments, but Vainshtein pointed out that the effect of nonlinearity might be important~\cite{vainshtein}.
In general relativity, the linear approximation is valid for distances much larger than the Schwarzschild radius of the source.
However, in massive gravity, linear approximation breaks down already at a distance much longer than the Schwarzschild radius.
Indeed, this Vainshtein effect has been shown to work in the DGP braneworld model~\cite{dvali,nicolis}.
Also, Rubakov~\cite{rubakov2} and Dubovsky~\cite{dubovsky} have proposed Lorentz violating massive gravity theories which evade pathologies related to the vDVZ discontinuity.
Therefore it is also important to put constraint on the graviton mass (or the graviton Compton wavelength $\lambda_g\equiv h/m_g c$).
When the graviton is massive, the gravitational potential is modified from Newtonian to Yukawa type.
Then, the effective gravitational constant depends on the distance from the gravitational source, which changes the Kepler's third law from Newtonian.
Verification of this third law in the solar system experiment puts a lower bound on the Compton wavelength as $\lambda_g>2.8 \times 10^{17}$cm~\cite{talmadge}.
 
In this paper, we estimate the possible constraint we can get on $\omega_{\mathrm{BD}}$ and $\lambda_g$ by detecting gravitational waves from the inspiral of precessing eccentric compact binaries with LISA~\cite{danzmann,lisa}.
The constraint on $\omega_{\mathrm{BD}}$ using binary gravitational waves was first discussed by Eardley~\cite{eardley}.
In Brans-Dicke theory, the additional scalar field contains the dipole radiation~\cite{will1977,zaglauer}.
This modifies the binary's orbital evolution from the one in general relativity.
Eardley mentioned that it is possible to constrain $\omega_{\mathrm{BD}}$ by measuring the rate of the secular decrease in the orbital period.
The change in the orbital evolution due to this dipole radiation modifies the phasing of the gravitational waveform.
Will~\cite{will1994} applied the matched filtering analysis and calculated how accurately one can determine the binary parameters including $\omega_{\mathrm{BD}}$ using advanced LIGO.
Scharre and Will~\cite{scharre} and Will and Yunes~\cite{yunes} did the same analysis using LISA.
Will and Yunes improved the previous works by showing the constraint dependences on LISA position noise, acceleration noise and arm lengths.
They also used slightly improved noise curve than the one used by Scharre and Will. 
However, these calculations do not include binary spins and also they used the pattern-averaged waveforms (having been averaged over the source direction $(\theta_{\mathrm{S}},\phi_{\mathrm{S}})$ and the direction of the orbital angular momentum $(\theta_{\mathrm{L}},\phi_{\mathrm{L}})$).
Berti \textit{et al.}~\cite{berti} calculated the determination errors including the effect of the spin-orbit coupling using LISA.
Also, they performed the Monte Carlo simulations in which they randomly distribute the directions and the orientations of $10^4$ sources over the sky, calculate the constraint on $\omega_{\mathrm{BD}}$ from each binary and take the average at the end.
According to their analysis, for a (1.4+1000)M$_{\odot}$ a neutron star (NS)/black hole (BH) binary of SNR=$\sqrt{200}$, 1 yr observation by LISA can put the bound $\omega_{\mathrm{BD}}>10799$ on average.

In massive graviton theories, the propagation speed of a gravitational wave depends on its frequency, which modifies the time of arrival from general relativity.
This also affects the phasing of the gravitational waveforms.
Will~\cite{will1998} included $\lambda_g$ into the binary parameters and carried out the matched filtering analysis, investigating how accurately one can determine $\lambda_g$ using advanced LIGO or LISA.
Will and Yunes~\cite{yunes} did the same analysis using the improved noise curve for LISA.  
Again, they did not include the effect of binary spins and also they used the pattern-averaged waveforms.
Berti \textit{et al.}~\cite{berti} estimated the constraint on $\lambda_g$ by performing the Monte Carlo simulations including the additional parameters of the spin-orbit coupling $\beta$ and the angles $\theta_{\mathrm{S}},\phi_{\mathrm{S}},\theta_{\mathrm{L}}$, and $\phi_{\mathrm{L}}$.
According to their analysis, for a ($10^6+10^6$)M$_{\odot}$ BH/BH binary at 3Gpc, 1 yr observation by LISA can put the bound $\lambda_g>1.33\times 10^{21}$cm on average.

In this paper, we improve the analysis of Ref.~\cite{berti} by taking the following effects into account: (i) the spin-spin coupling, (ii) the eccentricity, (iii) the spin precession.
(i) The spin-spin coupling $\sigma$ appears at the second post-Newtonian (2PN) order in the PN waveforms.
Berti \textit{et al.} reported that when they included both $\sigma$ and $\omega_{\mathrm{BD}}$ or $\sigma$ and $\lambda_g$ into parameters, they could not take the inverse of the Fisher matrix properly, because the ratio of the maximum eigenvalue to the minimum one of this matrix approaches the bound for the double precision computation.
Hence, we carried out our analysis in the quadruple precision.
(ii) As the binary radiates gravitational waves, it loses its energy and angular momentum~\cite{peters}.
Since this circularises the orbit as $e\propto f^{-19/18}$, usually the eccentricity is neglected in the analysis. 
We estimate how much the error of each parameter increases when we add eccentricity of 0.01 at 1 yr before coalescence into the binary parameters.
The reason for choosing this (seemingly rather small) fiducial value for eccentricity is explained in Sec~\ref{subsec-noprec}.
(iii) When the spin of the binary object is not zero, usually the spin-orbit effect and the spin-spin effect cause the spin vectors $\bm{S}_i$ (i=1,2) and the orbital angular momentum vector $\bm{L}$ to precess.
The parameter estimation errors including effects of precession in general relativity have been calculated by several authors~\cite{vecchio,lang, van1,van2,raymond,van3}.
Vecchio~\cite{vecchio} performed the Monte Carlo simulations of equal mass BH/BH binaries with LISA, taking only the leading spin-orbit terms in the precession equations.
In this case, the precession equations can be solved analytically up to some appropriate order.
This is called the \textit{simple precession approximation}~\cite{apostolatos}.
It has the property that $\bm{L}$ and $\bm{S}_i$ precess around a fixed vector $\bm{J}_0$ which is almost the same as the total angular momentum vector $\bm{J}$.
Another property is that the following three quantities, $\hat{\bm{L}}\cdot\hat{\bm{S}}$, $\hat{\bm{S}}_1\cdot\hat{\bm{S}}_2$ and $S$, become constant, where $\bm{S}=\bm{S}_1+\bm{S}_2$ is the total spin angular momentum and $S$ is its magnitude.
Lang and  Hughes~\cite{lang} solved the full precession equations, which include both spins of the binary stars, numerically and performed the Monte Carlo simulations of super massive black hole (SMBH)/SMBH binaries with LISA.
For $(10^7+10^6)M_{\odot}$ binaries at $z=1$, they found that the determination accuracy for $\Delta\mu/\mu$ increases by more than 2 orders of magnitude.
Van der Sluys \textit{et al.}~\cite{van1,van2,raymond} performed the Markov Chain Monte Carlo simulations for ground-based detectors, taking the simple precession approximation into account and performed the analysis that is beyond the Fisher matrix method.
In Ref.~\cite{van3}, they used the modeled gravitational wave signal injected into LIGO data so that the detector noise they used is more realistic than the Gaussian one.  
The precession gives some additional information to the waveform, which solves degeneracies among the parameters and enhances the parameter determination accuracy.
Here, we estimate how much the effect of precession enhances the determination accuracy in alternative theories of gravity.
We assume that the spin of one of the binary constituents can be neglected so that the analytic expressions in the simple precession approximation~\cite{vecchio} can be applied.  
The so-called Thomas precession is neglected in Ref.~\cite{vecchio} for simplicity but we found that this cannot be neglected for $\hat{\bm{L}}\cdot\hat{\bm{N}}\approx\pm 1$, where $\hat{\bm{N}}$ is the unit vector in the direction of the binary from the Sun.
For this reason, we approximately take this Thomas precession into account. 
Neglecting the spin of a NS is justified from observations~\cite{blanchet}.

Following the analysis of Berti \textit{et al.}~\cite{berti}, we assume that we detect gravitational waves from inspiral compact binaries from 1 yr before the coalescences with LISA.
We carry out the following Monte Carlo simulations. 
We distribute $10^4$ binaries over the sky and estimate the determination accuracy of binary parameters including $\omega_{\mathrm{BD}}$ and $\lambda_g$ for each binary using the Fisher matrix method.
We take the average at the end.
For the LISA noise curve, we use the same analytical approximation presented in Ref.~\cite{barack}.
    
We first performed our analysis using the pattern-averaged waveform. For the error estimation in Brans-Dicke theory, the inclusions of both $\sigma$ and eccentricity into parameters reduce the determination accuracy by an order of magnitude.
In particular, including eccentricity affects the estimation more than just including $\sigma$.
However, we found that if we impose the prior information on eccentricity, the constraint on $\omega_{\mathrm{BD}}$ remains almost as stringent as the one without including eccentricity into parameters.
For the analysis of massive graviton theories, the inclusion of these parameters only changes the results by a factor of a few.
In this case, the inclusion of $\sigma$ affects more than the inclusion of eccentricity.
Also, the prior distribution on eccentricity does not affect the result for the massive graviton theories.
Next, we performed the Monte Carlo simulations including precession. 
We found that in the Brans-Dicke case, the results are not so much affected by taking precession into account.
Using a NS/BH binary of $(1.4+1000)M_{\odot}$ with SNR=$\sqrt{200}$, estimation with all $\sigma$, eccentricity and precession taken into account leads to a constraint $\omega_{\mathrm{BD}}>2838$ on average. 
When we include the prior information on eccentricity, we expect that the constraint becomes $\omega_{\mathrm{BD}}>6944$, which is the same as the one without including eccentricity.
In the case of massive graviton theories, the inclusion of precession has a more remarkable effect.
Using a BH/BH binary of $(10^7+10^6)M_{\odot}$ at 3Gpc, estimation with all $\sigma$, eccentricity and precession taken into account can constrain $\lambda_g>3.06\times10^{21}$cm on average.
This constraint is 2.3 times stronger than the result obtained in Ref.~\cite{berti}.  
To compare our results with the ones obtained earlier using LISA~\cite{scharre,yunes,berti,will1998}, we list them in Table~\ref{table-comparison}.
Since all of these analyses do not take into account the errors coming from the use of approximate waveforms and the limitation of Fisher analysis~\cite{cutler-vallisneri}, the results shown in Table~\ref{table-comparison} are only approximate estimates.

\begin{table*}
\caption{\label{table-comparison}
Comparison of the constraints on $\omega_{\mathrm{BD}}$ and $\lambda_g$ in this paper and the ones obtained by several authors using LISA. Scharre and Will~\cite{scharre}, Will and Yunes~\cite{yunes}, and Will~\cite{will1998} performed the pattern-averaged analysis without including binary spins into parameters. 
Will and Yunes improved the previous works by showing the constraint dependences on LISA position noise, acceleration noise and arm lengths.
They also used slightly better noise curve than the one used by Scharre and Will.
Berti \textit{et al}.~\cite{berti} included spin-orbit coupling and performed Monte Carlo simulations.
We extended their analysis by taking spin-spin coupling, spin precession and eccentricity into account.
For the constraint on $\omega_{\mathrm{BD}}$, we show the one without eccentricity since we expect that the prior distribution $\Delta I_e >0$ reduces the results to the ones without including it. }
\begin{ruledtabular}
\begin{tabular}{c||ccccc}  
 & Will~\cite{will1998} & Scharre \& Will~\cite{scharre} & Will \& Yunes~\cite{yunes} & Berti \textit{et al}.~\cite{berti} & This paper \\
 & (1998) & (2002) & (2004) & (2005) & \\ \hline \hline
Binary mass & - & $(1.4+1000)M_{\odot}$ & $(1.4+1000)M_{\odot}$ & $(1.4+1000)M_{\odot}$ & $(1.4+1000)M_{\odot}$   \\ 
$\omega_{\mathrm{BD}}$  & - & 244549 & 203772 & 10799 & 6944  \\ \hline
Binary mass & $(10^7+10^6)M_{\odot}$ & - & $(10^6+10^6)M_{\odot}$ & $(10^6+10^6)M_{\odot}$ & $(10^7+10^6)M_{\odot}$   \\ 
$\lambda_g$ (cm)  & 6.9$\times$10$^{21}$ & - & 3.1$\times 10^{21}$ & 1.33$\times 10^{21}$ & 3.06$\times 10^{21}$   \\ 
\end{tabular}
\end{ruledtabular}
\end{table*}

The organisation of this paper is as follows.
In Sec.~\ref{sec-wave}, we review the waveforms of inspiral compact binaries in alternative theories of gravity.
We discuss how $df/dt$ is affected by the gravitational dipole radiation and the change in the propagation speed of gravitational waves.
In Sec.~\ref{sec-det}, we explain the output waveform of the detectors.
They are the superposition of the two polarised waves.
We consider the case of using two detectors and show the Fourier component of the restricted 2PN waveforms.
Section~\ref{sec-prec} discusses the effect of precession.
In Sec.~\ref{sec-simple}, we review how the angular momentum vectors $\bm{L}$ and $\bm{S}_i$ precess under the simple precession approximation.
In Sec.~\ref{sec-det-prec}, we discuss how the detector output expressed in Sec.~\ref{sec-det} changes when the orbital angular momentum $\bm{L}$ evolves with time.    
We also show how we treat the so-called \textit{Thomas precession}.
In Sec.~\ref{sec-par}, we describe how to estimate the determination errors of binary parameters using the matched filtering analysis.
In Sec.~\ref{sec-noise}, we present the noise curve of LISA that is needed when calculating the Fisher matrix.
In Sec.~\ref{sec-num}, we explain the setups and present the results of our numerical calculations.
In Sec.~\ref{sec-num-ave}, we use the pattern-averaged waveforms so that the angles $\theta_{\mathrm{S}},\phi_{\mathrm{S}},\theta_{\mathrm{L}},\phi_{\mathrm{L}}$ are not taken into parameters.
In Sec.~\ref{sec-num-mc}, we perform Monte Carlo simulations of $10^4$ binaries distributed over the sky. We calculate the determination errors for each binary and take the average.
In Sec.~\ref{sec-num-prec}, we carry out the same Monte Carlo simulations including the effect of the simple precession approximation. 
In Sec.~\ref{sec-conclusion}, we summarise our main results and discuss some future works.

\section{\label{sec-wave} BINARY GRAVITATIONAL WAVEFORMS IN ALTERNATIVE THEORIES OF GRAVITY}

\quad First, we calculate the gravitational waves coming from a binary system composed of compact objects in Brans-Dicke theory and massive graviton theories~\cite{berti}.
As the binary radiates gravitational waves, the orbit gets circularised.
Since we assume the observation starts 1 yr before coalescence, the eccentricity of this binary orbit is expected to be considerably small.
Therefore, we only include the leading term for the eccentricity in the phasing part.

In deriving the waveform, we use the post-Newtonian formalism, an expansion in terms of gravitational potential $U$ and $v^2$, where $v$ is the typical source velocity. 
The emitted gravitational waves are the superposition of harmonics at multiples of the orbital frequency.
The waveform $h(t)$ can be written schematically as 

\begin{equation}
h(t)=\mathrm{Re}\left( \sum_{n,m} h^n_m(t) e^{im\phi_{\mathrm{orb}}(t)} \right),
\end{equation}  
where $n$ labels PN order and $m$ is an index of the harmonics~\cite{flanagan}.
$\phi_{\mathrm{orb}}=\int^t\Omega(t')dt'$ is the orbital phase, where $\Omega(t)$ is the orbital angular velocity.
In this paper we use the ``restricted 2PN waveform.''
For the amplitude, we only take the leading Newtonian quadrupole term $h_2^0$, and for the phase part, we use $\phi_{\mathrm{orb}}(t)$ valid up to 2PN order.
This is because the correlation between two waveforms is much more sensitive to the phase information than to the amplitude when we perform the matched filtering analysis.
In this approximation the waveforms of + and $\times$ polarisations are

\begin{eqnarray}
h_{+}(t)&=&A_{+}\cos\phi(t), \\
h_{\times}(t)&=&A_{\times}\sin\phi(t),
\end{eqnarray} 
where $\phi(t)=2\phi_{\mathrm{orb}}(t)$ and 

\begin{eqnarray}
A_{+}&=&\frac{2m_1m_2}{rD_L}(1+(\hat{\bm{L}}\cdot\hat{\bm{N}})^2), \\
A_{\times}&=&-\frac{4m_1m_2}{rD_L}(\hat{\bm{L}}\cdot\hat{\bm{N}}).
\end{eqnarray}
Here $m_1$, $m_2$ are the two masses, $r$ is their orbital separation, $D_L$ is the luminosity distance between the source and observer, $\hat{\bm{L}}$ is the unit vector parallel to the orbital angular momentum, and $\hat{\bm{N}}$ is the unit vector pointing toward the source from the observer.
Please see Appendix~\ref{app-det-signal} for more explanations on polarisations.

Next, we need to calculate the phase $\phi(t)$.
First, we introduce the following useful mass parameters:

\begin{eqnarray}
M&\equiv&m_1+m_2, \\
\mu &\equiv& \frac{m_1m_2}{M}, \\
%\end{eqnarray}
%\\
%\begin{eqnarray}
\eta &\equiv&\frac{\mu}{M}, \\
\mathcal{M}&\equiv&\mu^{3/5}M^{2/5}=M\eta^{3/5}. 
\end{eqnarray}
%\begin{widetext}
They represent the total mass, the reduced mass, the symmetric mass ratio, and the chirp mass, respectively.
The rate at which the frequency changes because of the back-reaction due to the emission of gravitational waves is given by~\cite{berti,blanchet,cutlerharms}

\begin{widetext}
\begin{equation}
 \begin{split}
\frac{df}{dt}=\frac{96}{5}\pi^{8/3} \mathcal{M}^{5/3} f^{11/3} \biggl[ 1&+\frac{157}{24}I_e x^{-19/6}+\frac{5}{48}\mathcal{S}^2\bar{\omega }x^{-1}+\frac{96}{5}\beta_g \eta^{2/5} x-\left( \frac{743}{336}+\frac{11}{4}\eta \right) x \\
           &+(4\pi-\beta)x^{3/2}  
           +\left( \frac{34103}{18144}+\frac{13661}{2016}\eta +\frac{59}{18}\eta^2+\sigma \right) x^2  \biggr], \label{dfdt}
 \end{split}
\end{equation}
\end{widetext}
where we defined the squared typical velocity, $x\equiv v^2=(\pi M f)^{2/3}$.
The first term in the square brackets represents the lowest order quadrupole approximation of general relativity.
The second term is the contribution from small eccentricity~\cite{krolak}.
$I_e\equiv (\pi M)^{19/9}e_0^2f_0^{19/9}$ is the dimensionless asymptotic eccentricity invariant. (Note that our definition of $I_e$ is different from that in Ref.~\cite{krolak} by a factor $(\pi M)^{19/9}$ so that we set our $I_e$ dimensionless.)
$f_0$ is an arbitrary reference frequency and $e_0$ is the eccentricity when the frequency of quadrupole gravitational waves is $f_0$.
$I_e$ is constant up to the leading order in $e$.
The third term represents the dipole gravitational radiation in Brans-Dicke theory~\cite{will1994,scharre}. 
$\bar{\omega}\equiv \omega_{\mathrm{BD}}^{-1}$ is the inverse of the Brans-Dicke parameter.
$\mathcal{S}\equiv s_2-s_1$ with 

\begin{equation}
s_i\equiv -\left( \frac{\partial(\ln m_i)}{\partial (\ln G_{\mathrm{eff}})} \right)_0,
\end{equation}
which is called the \textit{sensitivity} of the $i$-th body.
Here, $G_{\mathrm{eff}}$ is the effective gravitational constant near the body and is proportional to the inverse of the Brans-Dicke scalar field there.
The subscript 0 denotes that we evaluate $s_i$ at a sufficiently large separation from the body.
This sensitivity roughly equals to the binding energy of the body per unit mass.
For example, the sensitivities of white dwarfs and neutron stars are $s_{\mathrm{WD}}\sim 10^{-3}$ and $s_{\mathrm{NS}}\sim 0.2$, respectively.
Because of the no hair theorem, black holes cannot have scalar charges and hence $s_{\mathrm{BH}}=0.5$.
From Eq.~(\ref{dfdt}), the contribution of dipole radiation becomes large as $\mathcal{S}$ increases.
Binaries with large $\mathcal{S}$ are the ones composed of bodies of different types.

The fourth term is the contribution from the mass of the graviton~\cite{will1998}.
When the graviton is massive, the propagation speed is slower than the speed of light, which modifies the gravitational wave phase from general relativity.
$\beta_g$ is given by~\cite{berti,will1998}

\begin{equation}
\beta_g \equiv \frac{\pi^2 D \mathcal{M}}{\lambda_g^2 (1+z)},
\end{equation}  
where $z$ is the cosmological redshift and $\lambda_g$ is the graviton Compton wavelength.
For a flat Universe ($\Omega_{\kappa}$=0, $\Omega_{\Lambda}+\Omega_{M}=1$), the distance $D$ is defined as

\begin{equation}
D\equiv\frac{1+z}{H_0}\int^z_0\frac{dz'}{(1+z')^2\sqrt{\Omega_{M}(1+z')^3+\Omega_{\Lambda}}}.
\end{equation} 
$H_0=72$km/s/Mpc is the Hubble constant, and $\Omega_{M}=0.3$ and $\Omega_{\Lambda}=0.7$ represent density parameters of matter and dark energy, respectively. 

The remaining terms are the usual higher order PN terms in general relativity.
The quantities $\beta$ and $\sigma$ represent spin-orbit and spin-spin contributions to the phase, respectively, given by

\begin{eqnarray}
\beta &\equiv &\frac{1}{12}\sum_{i=1}^{2}\chi_i\left( 113\frac{m_i^2}{M^2}+75\eta \right) \hat{\bm{L}}\cdot \hat{\bm{S}}_i,  \\
\sigma &\equiv&\frac{\eta}{48} \chi_1\chi_2 ( -247\hat{\bm{S}}_1 \cdot \hat{\bm{S}}_2
            +721(\hat{\bm{L}}\cdot \hat{\bm{S}}_1)( \hat{\bm{L}}\cdot \hat{\bm{S}}_2 )), \notag \\
\end{eqnarray}
where $\hat{\bm{S}}_i (i=1,2)$ are unit vectors in the direction of the spin angular momenta.
The spin angular momenta are given by $\bm{S}_i=\chi_i m_i^2 \hat{\bm{S}}_i$ where $\chi_i$ are the dimensionless spin parameters. 
For black holes, they must be smaller than unity, and for neutron stars, they are generally much smaller than unity.
It follows that $\lvert \beta\rvert \lesssim 9.4$ and $\lvert \sigma\rvert \lesssim 2.5$~\cite{berti}.

%\begin{widetext}
From $d\phi/dt=2\pi f$, we get the following time and phase evolution of the gravitational radiation by integrating Eq.~(\ref{dfdt}) with time,

\begin{widetext}
\begin{equation}
\begin{split}
t(f)=t_c-\frac{5}{256}\mathcal{M}(\pi \mathcal{M}f)^{-8/3} \biggl[1&-\frac{157}{43}I_e x^{-19/6} -\frac{\mathcal{S}^2}{12}\bar{\omega}x^{-1}
       -\frac{128}{5}\beta_g \eta^{2/5}x+\frac{4}{3}\left( \frac{743}{336}+\frac{11}{4}\eta \right) x  \\
       &-\frac{8}{5}(4\pi-\beta) x^{3/2} 
       +2\left( \frac{3058673}{1016064}+\frac{5429}{1008}\eta+\frac{617}{144}\eta^2 -\sigma \right) x^2 \biggr], \label{tf}
\end{split}
\end{equation}

\begin{equation}
\begin{split}
\phi(f)=\phi_c-\frac{1}{16}(\pi\mathcal{M}f)^{-5/3}\biggl[1&-\frac{785}{344}I_e x^{-19/6}-\frac{25}{336}\mathcal{S}^2\bar{\omega}x^{-1}
        -32\beta_g \eta^{2/5}x+\frac{5}{3}\left( \frac{743}{336}+\frac{11}{4}\eta \right) x  \\
       &-\frac{5}{2}(4\pi-\beta) x^{3/2} 
       +5\left( \frac{3058673}{1016064}+\frac{5429}{1008}\eta+\frac{617}{144}\eta^2 -\sigma \right) x^2 \biggr],
\end{split}
\end{equation}
%\end{widetext}
where $t_c$ and $\phi_c$ are the time and phase at coalescence.
\end{widetext}

\section{\label{sec-det}DETECTOR RESPONSE}

\begin{figure}[thbp]
% \begin{center}
  \includegraphics[scale=.4,clip]{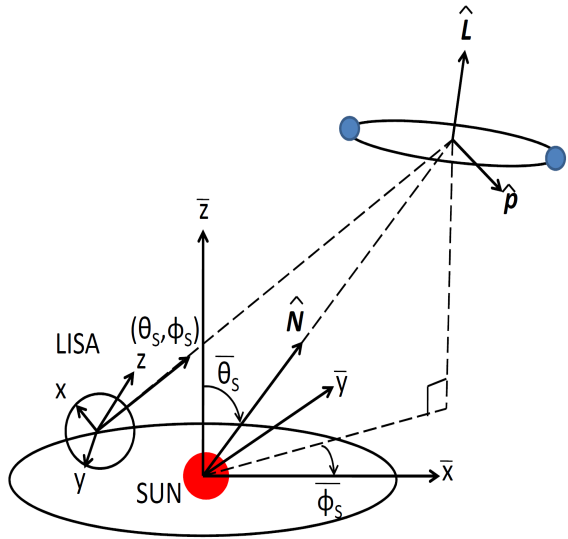} 
%   \end{center}
 \caption{\label{fig1} We use two types of coordinates: (i)a barred barycentric frame $(\bar{x},\bar{y},\bar{z})$ tied to the ecliptic and centred in the solar system barycentre, (ii) an unbarred detector frame $(x,y,z)$, centred in the barycentre of the triangle and attached to the detector.}
\end{figure}

\quad LISA is an all-sky monitor with a quadrupolar antenna pattern and consists of three drag-free spacecrafts arranged in an equilateral triangle, 5$\times$ $10^6$km apart.
Each spacecraft contains a free-falling mirror so that LISA forms three-arm interferometers with opening angles 60$^\circ$.
The barycentre of each triangle orbits the Sun 20$^\circ$ behind the Earth and the detector plane is tilted by 60$^\circ$ with respect to the ecliptic.

Following Ref.~\cite{cutler1998}, we introduce two Cartesian reference frames: (i) a barred barycentric frame $(\bar{x},\bar{y},\bar{z})$ tied to the ecliptic and centred in the solar system barycentre, with $\hat{\bar{\bm{z}}}$ (unit vector in ${\bar{\bm{z}}}$ direction) normal to the ecliptic and $\bar{x}\bar{y}$-plane aligned with the ecliptic, (ii) an unbarred detector frame $(x,y,z)$, centred in the barycentre of the triangle and attached to the detector, with $\hat{\bm{z}}$ normal to the detector plane (see Fig.~\ref{fig1}). The orbit of the detector barycentre can be written as

\begin{equation}
\bar{\theta}(t)=\pi/2, \qquad \bar{\phi}(t)=2\pi t/T,
\end{equation}
where $T=1$year and we have assumed $\bar{\phi}(t=0)=0$.

The interferometer having three arms corresponds to having two individual detectors.
We label each detector as detector $\mathrm{I}$ and $\mathrm{II}$, respectively.
The waveforms measured by each detector are given as

\begin{equation}
%\begin{split}
h_{\alpha}(t)=\frac{\sqrt{3}}{2} \frac{2m_1m_2}{rD_L} 
                   A_{\mathrm{pol},\alpha}(t) \cos \left[ \phi(t) + \varphi_{\mathrm{pol},\alpha}(t)+\varphi_{D}(t) \right], 
                               \label{waveform1}
%\end{split}
\end{equation}
where $\alpha=\mathrm{I}, \mathrm{II}$ represents the detector number (see Appendix~\ref{app-det-signal} for more details).
The polarisation amplitude $A_{\mathrm{pol},\alpha}(t)$, the polarisation phases $\varphi_{\mathrm{pol},\alpha}(t)$ and the Doppler phase $\varphi_{D}(t)$ are defined as 

\begin{widetext}
\begin{eqnarray}
A_{\mathrm{pol},\alpha}(t)&=&\sqrt{(1+(\hat{\bm{L}}\cdot\hat{\bm{N}})^2)^2F_{\alpha}^{+}(t)^2+4(\hat{\bm{L}}\cdot\hat{\bm{N}})^2F_{\alpha}^{\times}(t)^2}, \label{Apol} \\
 \cos(\varphi_{\mathrm{pol},\alpha}(t))&=&\frac{(1+(\hat{\bm{L}}\cdot\hat{\bm{N}})^2)F^{+}_{\alpha}(t)}{A_{\mathrm{pol},\alpha}(t)}, \\
 \sin(\varphi_{\mathrm{pol},\alpha}(t))&=&\frac{2(\hat{\bm{L}}\cdot\hat{\bm{N}}) F^{\times}_{\alpha}(t)}{A_{\mathrm{pol},\alpha}(t)}, \label{phipol} \\
\varphi_{D}(t)&=&2\pi f(t) R \sin \bar{\theta}_{\mathrm{S}} \cos[\bar{\phi}(t)-\bar{\phi}_{\mathrm{S}}], \label{doppler-phase}
\end{eqnarray}
\end{widetext}
where $\hat{\bm{L}}$ is the unit vector parallel to the orbital angular momentum and $\hat{\bm{N}}$ is the unit vector pointing toward the centre of mass of the binary, and $F_{\alpha}^{+}$ and $F_{\alpha}^{\times}$ are the beam-pattern functions of $+$ and $\times$ polarisation modes for each detector, shown in Appendix~\ref{app-det-signal}. 
$(\bar{\theta}_{\mathrm{S}},\bar{\phi}_{\mathrm{S}})$ represents the direction of the source in the solar barycentre frame and $R$ represents 1 AU. 
The Doppler phase denotes the difference between the phase of the wave front at the detector and the phase of the wavefront at the solar system barycentre.
It arises from the motion of the detector around the Sun.
The factor $\sqrt{3}/2$ in Eq.~(\ref{waveform1}) comes from the 60$^\circ$ opening angle of adjacent detector arms of LISA.

Later, we estimate the accuracy of determination of the binary parameters using the matched filtering analysis, where we work on the Fourier domain.
Therefore we calculate the Fourier transform of the signal,

\begin{equation}
\tilde{h}_{\alpha}(f)=\int^{\infty}_{-\infty} dt\, e^{2\pi ift} h_{\alpha}(t).
\end{equation}
To evaluate this quantity, we use the stationary phase approximation~\cite{flanagan}.
When conditions $d \ln A/dt \ll d \phi/dt$ and $d^2 \phi/dt^2 \ll (d \phi /dt)^2$ are satisfied, the saddle point method can be used and the Fourier transform of a function $B(t)=A(t) \cos \phi (t)$ becomes

\begin{equation}
\tilde{B}(f) \approx \frac{1}{2} A(t) \left( \frac{df}{dt} \right)^{-1/2} \exp{i(2\pi ft(f)-\phi(f)-\pi/4)}.
\end{equation}
Under this approximation, the Fourier component of the waveform $\tilde{h}(f)$ becomes~\cite{berti}

\begin{equation}
\tilde{h}(f)=\frac{\sqrt{3}}{2}\mathcal{A}f^{-7/6}e^{i\Psi (f)} \left[ \frac{5}{4}A_{\mathrm{pol},\alpha}(t(f)) \right] e^{-i \left( \varphi_{\mathrm{pol},\alpha}+\varphi_D \right)}, \label{waveform}
\end{equation}
where the amplitude $\mathcal{A}$ and the phase $\Psi(f)$ are given by

\begin{equation}
\mathcal{A}=\frac{1}{\sqrt{30}\pi^{2/3}}\frac{\mathcal{M}^{5/6}}{D_L},  \label{amp-noangle}
\end{equation}

\begin{widetext}
\begin{equation}
 \begin{split}
\Psi(f)= &2\pi ft_c-\phi_c -\frac{\pi}{4}+\frac{3}{128}(\pi \mathcal{M}f)^{-5/3} \biggl[1-\frac{5}{84}\mathcal{S}^2\bar{\omega}x^{-1} 
                 -\frac{128}{3}\beta_g\eta^{2/5}x-\frac{2355}{1462}I_e x^{-19/6} \\
                 &+ \left( \frac{3715}{756}+\frac{55}{9}\eta \right)x 
                  -4(4\pi-\beta)x^{3/2} 
           + \left( \frac{15293365}{508032}+\frac{27145}{504}\eta+\frac{3085}{72}\eta^2-10\sigma \right) x^2  \biggr]. \label{Psi-noangle}
 \end{split}
\end{equation}
\end{widetext}
Also, when we integrate out the angle dependence from the waveform~(\ref{waveform}), it becomes

\begin{equation}
\tilde{h}(f)=\frac{\sqrt{3}}{2}\mathcal{A}f^{-7/6}e^{i\Psi (f)}. \label{wave-noangle}
\end{equation}

\section{\label{sec-prec}PRECESSION}

\quad In this section, we introduce an additional effect, the precession.
The spin-orbit interaction and spin-spin interaction change the orientations of the orbital angular momentum vector $\bm{L}$ and the spin vectors $\bm{S}_i$.
These vectors precess over a time scale longer than the orbital period but shorter than the observation time scale.
This effect drastically changes the detected waveforms.

\subsection{\label{sec-simple}Simple Precession}

In this paper, we assume that one of the spins of the binary constituents is negligible (i.e. $\bm{S}_1\sim 0$).
%We also assume that the orbital angular momentum $\bm{L}$ is neither parallel nor antiparallel to the total spin angular momentum $\bm{S}$($=\bm{S}_1+\bm{S}_2$).
Then, the precession equations become

\begin{eqnarray}
\dot{L}&=&-\frac{32}{5}\frac{\mu^2}{r}\left( \frac{M}{r} \right)^{5/2} \label{40a}, \\
\dot{S}&=&0, \\
\dot{\hat{\bm{L}}}&=&\left( 2+\frac{3}{2}\frac{m_2}{m_1}\right)\frac{\bm{J}}{r^3}\times \hat{\bm{L}}, \label{40c} \\
\dot{\hat{\bm{S}}}&=&\left( 2+\frac{3}{2}\frac{m_2}{m_1}\right)\frac{\bm{J}}{r^3}\times \hat{\bm{S}}, \label{40d} 
\end{eqnarray}
where $\bm{J}$ is the total angular momentum $\bm{J}\equiv \bm{L}+\bm{S}$.
Under this simple precession approximation, the following quantities become constant during the inspiral, $\bm{S}_1\cdot\bm{S}_2$, $\kappa\equiv\hat{\bm{L}}\cdot\hat{\bm{S}}$, and the magnitude of the total spin angular momentum $S\equiv\lvert \bm{S}_1+\bm{S}_2 \rvert$.
From the above equations, it can be seen that both $\hat{\bm{L}}$ and $\hat{\bm{S}}$ precess around $\bm{J}$ with angular velocity 

\begin{equation}
\Omega_p=\left( 2+\frac{3}{2}\frac{m_2}{m_1}\right)\frac{J}{r^3}. \label{omega-precess}
\end{equation}
In general, the precessing time scale $\Omega_p^{-1}$ is shorter than the inspiral time scale $L/\lvert \dot{L}\rvert$.
Therefore $\bm{J}$ changes in magnitude but the direction is almost constant. 
Then, the analytic form of $\hat{\bm{L}}$ can be obtained up to some approximate orders (see Appendix~\ref{app-prec}).

\subsection{\label{sec-det-prec}Detector Response}

When precession is taken into account, the principal axis $\hat{\bm{N}}\times\hat{\bm{L}}$ varies with time so that the GW phase $\phi(t)$ no longer equals to twice the orbital phase $\phi_{\mathrm{orb}}(t)\equiv\int\Omega(t) dt$.
We define this difference 2$\delta\phi(t)$ by

\begin{equation}
\phi(t) =2\phi_{\mathrm{orb}}(t)+2\delta\phi(t). \label{phi-prec}
\end{equation}
This $\delta\phi(t)$ is the so-called \textit{Thomas precession phase}.
We specify the constant of integration so that $\phi_{\mathrm{orb}}(t_c)=\phi_c$.
In general, $\delta\phi(t_c)\neq 0$ so that $\phi_c$ does not equal to $\phi(t_c)$ anymore in this case. 
From Eqs.~(\ref{output2}) and~(\ref{phi-prec}), the detector output $h_{\alpha}(t)$ becomes

\begin{widetext}
\begin{eqnarray}
h_{\alpha}(t)&=&\frac{\sqrt{3}}{2} A_{+} F_{\alpha}^+(\theta_{\mathrm{S}},\phi_{\mathrm{S}},\psi_{\mathrm{S}}) 
                      \cos2 (\phi_{\mathrm{orb}}(t)+\delta\phi(t))
                     +\frac{\sqrt{3}}{2} A_{\times} F_{\alpha}^{\times}(\theta_{\mathrm{S}},\phi_{\mathrm{S}},\psi_{\mathrm{S}}) 
                      \sin2 (\phi_{\mathrm{orb}}(t)+\delta\phi(t)) \notag \\ 
                 &=&\frac{\sqrt{3}}{2} \frac{2m_1m_2}{rD}(F^{\cos}_{\alpha}\cos 2\phi_{\mathrm{orb}}+F^{\sin}_{\alpha}\sin 2\phi_{\mathrm{orb}}), \label{out-prec}
\end{eqnarray}
%\end{widetext}
where $\alpha=\mathrm{I}, \mathrm{II}$ labels the detector number and $F^{\cos}_{\alpha}$ and $F^{\sin}_{\alpha}$ are defined as follows:

%\begin{widetext}
\begin{eqnarray}
F^{\cos}_{\alpha}&\equiv &(1+(\hat{\bm{L}}\cdot\hat{\bm{N}})^2)F_{\alpha}^+\cos 2\delta\phi
                             -2(\hat{\bm{L}}\cdot\hat{\bm{N}})F_{\alpha}^{\times}\sin 2\delta\phi, \\
F^{\sin}_{\alpha}&\equiv &-(1+(\hat{\bm{L}}\cdot\hat{\bm{N}})^2)F_{\alpha}^+\sin 2\delta\phi
                             -2(\hat{\bm{L}}\cdot\hat{\bm{N}})F_{\alpha}^{\times}\cos 2\delta\phi.
\end{eqnarray} 
\end{widetext}
Following Eq.~(\ref{waveform1}), we express this output~(\ref{out-prec}) in terms of an amplitude and phase form, and we also take the motions of the detectors into account to obtain

\begin{equation}
h_{\alpha}(t)=\frac{\sqrt{3}}{2} \frac{2m_1m_2}{rD}A_{\mathrm{pol},\alpha}^{\mathrm{prec}}(t)
                     \cos(2\phi_{\mathrm{orb}}(t)+\varphi_{\mathrm{pol},\alpha}^{\mathrm{prec}}(t)+\varphi_{D}(t)),
\end{equation}
where $A_{\mathrm{pol},\alpha}^{\mathrm{prec}}(t)$ and $\varphi_{\mathrm{pol},\alpha}^{\mathrm{prec}}(t)$ are given as

\begin{eqnarray}
A_{\mathrm{pol},\alpha}^{\mathrm{prec}}(t)&=&\sqrt{(F^{\cos}_{\alpha}(t))^2+(F^{\sin}_{\alpha}(t))^2}, \\
\cos(\varphi_{\mathrm{pol},\alpha}^{\mathrm{prec}}(t))&=&\frac{F^{\cos}_{\alpha}(t)}{A_{\mathrm{pol},\alpha}^{\mathrm{prec}}(t)}, \\
\sin(\varphi_{\mathrm{pol},\alpha}^{\mathrm{prec}}(t))&=&-\frac{F^{\sin}_{\alpha}(t)}{A_{\mathrm{pol},\alpha}^{\mathrm{prec}}(t)}. 
\end{eqnarray}

The quantities $(\hat{\bm{L}}\cdot\hat{\bm{N}})$,$(\hat{\bm{L}}\cdot\hat{\bm{z}})$ and $[\hat{\bm{N}}\cdot(\hat{\bm{L}}\times\hat{\bm{z}})]$, which are needed to compute the polarisation angle $\psi_{\mathrm{S}}(t)$ in the beam-pattern coefficients $F_{\alpha}^+$ and $F_{\alpha}^{\times}$, are expressed in Appendix~\ref{app-prec}.

Finally, we need to calculate the Thomas precession phase $\delta\phi(t)$.
Apostolatos \textit{et al.}~\cite{apostolatos} derived an explicit form as

\begin{equation}
\delta\phi(t)=-\int^{tc}_{t}dt \, \left( \frac{\hat{\bm{L}}\cdot\hat{\bm{N}}}{1-(\hat{\bm{L}}\cdot\hat{\bm{N}})^2} \right)
                            (\hat{\bm{L}}\times\hat{\bm{N}})\cdot\dot{\hat{\bm{L}}}. \label{delta-phi}
\end{equation}
This expression includes an integration which makes the computational time very long. 
Vecchio~\cite{vecchio} estimated the binary parameter accuracies by choosing a few random sources with and without including $\delta\phi(t)$ and found that this term did not affect the results, and concluded that this term could be neglected.
This is true for the binaries for which $\hat{\bm{L}}\cdot\hat{\bm{N}}$ never becomes close to $\pm1$.
However, when $\hat{\bm{L}}\cdot\hat{\bm{N}}\approx \pm1$, the direction of the principal axis ($\hat{\bm{N}}\times\hat{\bm{L}}$) changes rapidly with time, and hence the polarisation angle $\psi_{\mathrm{S}}(t)$ also changes rapidly.
Since it is the Thomas precession phase $\delta\phi(t)$ that cancels this rapid change, $\delta\phi(t)$ cannot be neglected in this case.

\begin{figure}[t]
 \begin{center}
  \includegraphics[scale=.4,clip]{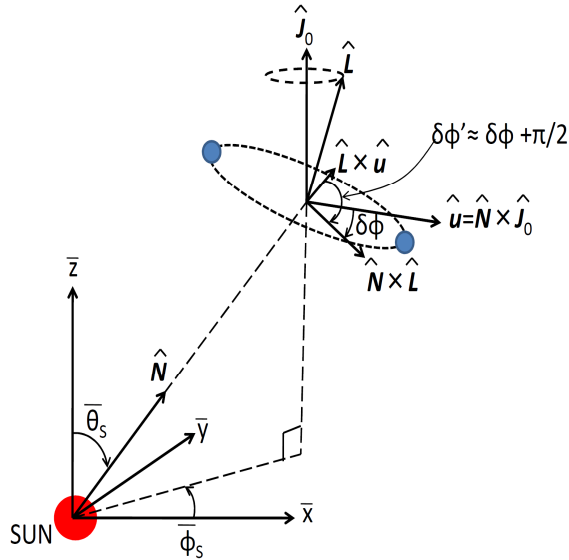} 
   \end{center}
 \caption{The Thomas precession phase $\delta\phi(t)$ is the angle from the vector $\hat{\bm{u}}$ to the principal axis $\hat{\bm{N}}\times\hat{\bm{L}}$. We define $\delta\phi '$ as the angle from the vector $\hat{\bm{L}}\times\hat{\bm{u}}$ to the one $\hat{\bm{N}}\times\hat{\bm{L}}$. Notice that these two vectors always lie on the orbital plane. $\delta\phi '$ equals to $\delta\phi + \pi /2$ up to $O(\lvert \delta\bm{L} \bm{} \rvert/L)$.}
\label{fig3}
\end{figure}

The direction which $\delta\phi$ is measured from is arbitrary.
It can be seen from Eq.~(\ref{delta-phi}) that Apostolatos \textit{et al.}~\cite{apostolatos} defined $\delta\phi$ as the angle measured from the principal axis at the time of coalescence.
In general, since $(\hat{\bm{N}}\times\hat{\bm{L}})_{t=t_c}$ does not lie on the orbital plane at a time $t$, we need to follow the evolution of the principal axis to calculate the Thomas precession $\delta\phi(t)$.
This is the reason why the integration appears in Eq.~(\ref{delta-phi}). 

Here, we try to derive an approximate expression of $\delta\phi(t)$ that is not in the integral form.
To do so, we use a specific vector $\hat{\bm{L}}\times\hat{\bm{u}}$ that always lies on the orbital plane, where $\hat{\bm{u}}$ is a constant unit vector.
In the limit of a large separation (i.e. $t\rightarrow -\infty$ or $L\rightarrow\infty$), $\hat{\bm{L}}$ approaches to $\hat{\bm{J}}_0$.
Using this $\hat{\bm{J}}_0$, we take $\hat{\bm{u}}=\hat{\bm{N}}\times\hat{\bm{J}}_0$ (see Fig.~\ref{fig3}).
We define an approximate Thomas precession phase $\delta\phi(t)$ by the angle from this vector $\hat{\bm{u}}$ to the principal axis $\hat{\bm{N}}\times\hat{\bm{L}}$. 
(We may take the vector $\hat{\bm{u}}$ to be the principal axis at the time of coalescence $(\hat{\bm{N}}\times\hat{\bm{L}})_{t=t_c}$. Then, $\delta\phi(t_c)$ becomes 0 and $\phi(t_c)=\phi_c$. However, the computation is easier for choosing $\hat{\bm{u}}$ to be $\hat{\bm{N}}\times\hat{\bm{J}}_0$.)
We denote the angle from $\hat{\bm{L}}\times\hat{\bm{u}}$ to $\hat{\bm{N}}\times\hat{\bm{L}}$ as $\delta\phi '$, and the difference between $\bm{J}_0$ and $\bm{L}$ as $\delta\bm{L}$; $\bm{L}=\bm{J}_0+\delta\bm{L}$.  
Then, $\delta\phi '$ can be related to $\delta\phi$ as $\delta\phi '=\delta\phi+\pi/2$ up to $O(\lvert \delta\bm{L} \bm{} \rvert/L)$.
As $\delta\phi '$ always lie on the orbital plane, by using this angle we can express $\delta\phi(t)$ without integration.
$\sin\delta\phi$ and $\cos\delta\phi$ can be written as 

\begin{eqnarray}
\sin\delta\phi &\approx &-\cos\delta\phi ' \notag \\
                      &=&-\frac{(\hat{\bm{L}}\times\hat{\bm{N}})\cdot(\hat{\bm{L}}\times\hat{\bm{u}})}
                       {\lvert \hat{\bm{L}}\times\hat{\bm{N}} \rvert \lvert \hat{\bm{L}}\times\hat{\bm{u}} \rvert}  \notag \\
                      &=&\frac{(\hat{\bm{L}}\cdot\hat{\bm{N}})(\hat{\bm{L}}\cdot\hat{\bm{u}})}
                       {\sqrt{(1-(\hat{\bm{L}}\cdot\hat{\bm{N}})^2)(1-(\hat{\bm{L}}\cdot\hat{\bm{u}})^2)}}, \\
\cos\delta\phi &\approx &\sin\delta\phi ' \notag \\
                       &=&-\frac{\hat{\bm{N}}\cdot(\hat{\bm{L}}\times\hat{\bm{u}})}
                        {\sqrt{(1-(\hat{\bm{L}}\cdot\hat{\bm{N}})^2)(1-(\hat{\bm{L}}\cdot\hat{\bm{u}})^2)}}.
\end{eqnarray}
Therefore the Thomas precession phase $\delta\phi(t)$ can be expressed as

\begin{equation}
\delta\phi\approx \arctan \left( -\frac{(\hat{\bm{L}}\cdot\hat{\bm{N}})(\hat{\bm{L}}\cdot\hat{\bm{u}})}
                  {\hat{\bm{N}}\cdot(\hat{\bm{L}}\times\hat{\bm{u}})} \right). \label{thomas-new}
\end{equation}
The explicit forms of $\hat{\bm{L}}\cdot\hat{\bm{N}}$, $\hat{\bm{L}}\cdot\hat{\bm{u}}$ and $\hat{\bm{N}}\cdot(\hat{\bm{L}}\times\hat{\bm{u}})$ are given in Appendix~\ref{app-prec}.

\section{\label{sec-par}PARAMETER ESTIMATION}

\quad We use the matched filtering analysis to estimate the determination errors of the binary parameters $\bm{\theta}$~\cite{finn,flanagan}. 
We assume that the detector noise is stationary and Gaussian.
``Stationary" means that different Fourier components $\tilde{n}(f)$ of the noise are uncorrelated and we have

\begin{equation}
\left\langle \tilde{n}^{*}(f)\tilde{n}(f')\right\rangle =\delta(f-f')\frac{1}{2}S_n(f). \label{stationary}
\end{equation}
Here, $\left\langle\dots\right\rangle$ denotes the expectation value and $S_n(f)$ is the noise spectral density. 
%It can be seen that the noise r.m.s. is proportional to $\frac{1}{2}S_n(f)$.
Then the noise takes the Gaussian probability distribution given by 

\begin{equation}
\begin{split}
p(n_0) &\propto \exp\left[ -\frac{1}{2}\int^{\infty}_{-\infty}df \, \frac{\lvert \tilde{n}_0(f) \rvert^2}{(1/2)S_n(f)} \right] \\
         &\propto \exp \left[-\frac{1}{2}(n_0|n_0)\right], \label{gauss}
\end{split}
\end{equation}
where we have defined the inner product as 

\begin{equation}
\begin{split}
(A|B)&=\mathrm{Re}\int ^{\infty}_{-\infty}df \, \frac{\tilde{A}^{*}(f)\tilde{B}(f)}{\frac{1}{2}S_n(f)} \\
       &=4 \mathrm{Re}\int ^{\infty}_{0}df \, \frac{\tilde{A}^{*}(f)\tilde{B}(f)}{S_n(f)}. \label{scalar-prod}
\end{split}
\end{equation}
The signal to noise ratio(SNR) for a given $h$ is  

\begin{equation}
\rho[h]\equiv \sqrt{(h|h)}.
\end{equation}

The detected signal $s(t)=h(t,\bm{\theta}_t)+n_0(t)$ is the sum of the gravitational wave signal $h(t;\bm{\theta}_t)$ and the noise $n(t)$, 
 where $\bm{\theta}_t$ is the true binary parameters and $n_0(t)$ is the noise of this specific measurement. Then, Eq.~(\ref{gauss}) becomes 

\begin{equation}
p(\theta_t|s)\propto p^{(0)}(\theta_t)\exp\left[ (h_t|s)-\frac{1}{2}(h_t|h_t) \right], \label{prob}
\end{equation}
where $p^{(0)}(\theta_t)$ represents the distribution of prior information and $h_t\equiv h(\bm{\theta}_t) $.
We determine the binary parameters as $\hat{\bm{\theta}}$ that maximise the probability distribution $p(\theta_t|s)$. 
Then $\hat{\bm{\theta}}$ is the solution of the following equation,

\begin{equation}
(\partial_ih_t|s)-(\partial_ih_t|h_t)=0,
\end{equation}
where $\partial_i \equiv \frac{\partial}{\partial\theta^i_t}$.
We denote the error in the determination of $\theta^i$ as $\Delta\theta^i$; $\theta^i=\hat{\theta}^i+\Delta\theta^i$.
Next, we expand Eq.~(\ref{prob}) in powers of $\Delta\theta^i$ up to quadratic order and we get

\begin{equation}
p(\bm{\theta}|s)\propto p^{(0)}(\bm{\theta})\exp\left[ -\frac{1}{2}\Gamma_{ij}\Delta\theta^i\Delta\theta^j \right],
\end{equation}
where $\Gamma_{ij}=(\partial_i\partial_jh|h-s)+(\partial_ih|\partial_jh)$ is called Fisher matrix.
Since $h-s=-n$, in the limit of large SNR, we can neglect the first term of $\Gamma_{ij}$ and 

\begin{equation}
\Gamma_{ij}=(\partial_ih|\partial_jh).
\end{equation} 
When we use both detectors $\mathrm{I}$ and $\mathrm{II}$, the Fisher matrix becomes

\begin{equation}
\Gamma_{ij}=(\partial_ih_{\mathrm{I}}|\partial_jh_{\mathrm{I}})+(\partial_ih_{\mathrm{II}}|\partial_jh_{\mathrm{II}}).
\end{equation}
We take into account our prior information on the maximum spin by assuming

\begin{equation}
p^{(0)}(\bm{\theta})\propto \exp\left[ -\frac{1}{2}\left( \frac{\beta}{9.4} \right)^2- \frac{1}{2}\left( \frac{\sigma}{2.5} \right)^2\right]. \label{prior}
\end{equation}
Then, the rms of $\Delta\theta^i$ can be calculated by taking the square root of the diagonal elements of the covariance matrix $\Sigma^{ij}$, which is the inverse of the Fisher matrix:

\begin{equation}
\Sigma^{ij}\equiv \left\langle\Delta\theta^i\Delta\theta^j\right\rangle =(\tilde{\Gamma}^{-1})^{ij},
\end{equation}
and $\tilde{\Gamma}_{ij}$ is defined by

\begin{equation}
p^{(0)}(\bm{\theta})\exp\left[ -\frac{1}{2}\Gamma_{ij}\Delta\theta^i\Delta\theta^j \right] \equiv \exp\left[ -\frac{1}{2}\tilde{\Gamma}_{ij}\Delta\theta^i\Delta\theta^j \right]
\end{equation}

\section{\label{sec-noise}LISA NOISE SPECTRUM}

\quad In this section, we introduce the noise spectrum of LISA following Ref.~\cite{barack}.
The instrumental noise spectral density for LISA is given as

\begin{widetext}
\begin{equation}
S_h^{\mathrm{inst}}(f)=\left[ 9.18\times 10^{-52}\left( \frac{f}{1\ \mathrm{Hz}} \right)^{-4}+1.59\times 10^{-41}
                                                        +9.18\times 10^{-38} \left( \frac{f}{1\mathrm{Hz}} \right)^2 \right] \ \mathrm{Hz^{-1}}.
\end{equation}
\end{widetext}

\begin{figure}[htbp]
  \includegraphics[scale=.5,clip]{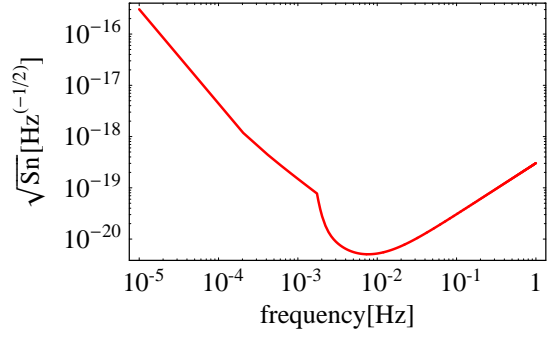} 
 \caption{\label{noise-lisa} The noise spectral density for LISA.}
\end{figure}

Besides instrumental noise, there are foreground confusion noises.
These confusion noise spectral densities and the energy densities of gravitational waves are related as 

\begin{equation}
S_h^{\mathrm{conf}}=\frac{3}{5\pi}f^{-3}\rho_c\Omega_{\mathrm{GW}}, \label{confusion}
\end{equation}
where $\rho_c\equiv \frac{3H_0^2}{8\pi}$ is the critical energy density of the Universe and 

\begin{equation}
\Omega_{\mathrm{GW}}\equiv\frac{1}{\rho_c}\frac{d \rho_{\mathrm{GW}}}{d \ln f}
\end{equation}
is the energy density of gravitational waves per log frequency normalised by $\rho_c$.
The energy density of gravitational waves coming from extra-galactic white dwarf binaries has been estimated as
$\Omega_{\mathrm{GW}}=3.6\times10^{-12}(f/10^{-3}\mathrm{Hz})^{2/3}$ and the noise spectral density becomes~\cite{farmer}

\begin{equation}
S_h^{\mathrm{ex-gal}}(f)=4.2\times10^{-47}\left(\frac{f}{1\ \mathrm{Hz}}\right)^{-7/3} \ \mathrm{Hz^{-1}}.
\end{equation}
On the other hand, the energy density of gravitational waves coming from galactic white dwarf binaries has been calculated as 50 times larger than the one coming from extra-galactic white dwarf binaries.
Therefore the noise spectral density becomes~\cite{nelemans}

\begin{equation}
S_h^{\mathrm{gal}}(f)=2.1\times10^{-45}\left(\frac{f}{1\ \mathrm{Hz}}\right)^{-7/3} \ \mathrm{Hz^{-1}}.
\end{equation}
We compute the total noise spectral density as

\begin{widetext}
\begin{equation}
S_h(f)=\min\left[ \frac{S_h^{\mathrm{inst}}(f)}{\exp(-\kappa T^{-1}dN/df)},\ 
            S_h^{\mathrm{inst}}(f)+S_h^{\mathrm{gal}}(f) \right] +S_h^{\mathrm{ex-gal}}(f). \label{noise-LISA}
\end{equation}
\end{widetext}
Here $dN/df$ is the number density of galactic white dwarf binaries per unit frequency, for which we use the estimate given in Ref.~\cite{hughes}

\begin{equation}
\frac{dN}{df}=2\times10^{-3} \mathrm{Hz}^{-1}\left(\frac{f}{1 \ \mathrm{Hz}} \right)^{-11/3}.
\end{equation}
$\kappa\simeq 4.5$ is the average number of frequency bins that are lost when each galactic binary is fitted out.
This noise curve is drawn in Fig.~\ref{noise-lisa}

\section{NUMERICAL CALCULATIONS AND RESULTS}
\label{sec-num}

\quad Following Berti \textit{et al.}~\cite{berti}, we estimate how accurately we can determine binary parameters by detecting inspiral gravitational waves with LISA.
The total noise spectral density for LISA is given by Eq.~(\ref{noise-LISA}) and we introduce cutoff frequencies for LISA as follows:

\begin{equation}
f_{\mathrm{low}}=10^{-5} \mathrm{Hz}, \qquad f_{\mathrm{high}}=1 \mathrm{Hz}.
\end{equation}
Here, we use the matched filtering analysis, assuming that the observation starts 1 yr before the coalescence.
We assume that we use two detectors which corresponds to one triangle interferometer as mentioned in Sec.~\ref{sec-det}.
We numerically calculate the determination error of each parameter which is given by the square root of the corresponding diagonal element of $\Sigma^{ij}$. 
We use the Gauss-Jordan elimination~\cite{numerical recipes} for inverting the Fisher matrix $\tilde{\Gamma}_{ij}$ to obtain $\Sigma^{ij}$. 
In calculating this $\tilde{\Gamma}_{ij}$, we need to perform the integration in the frequency domain.
Following Ref.~\cite{berti}, we take the frequency range of this integration as $(f_{\mathrm{in}},f_{\mathrm{fin}})$;

\begin{equation}
f_{\mathrm{in}}=\max \bigl\{ f_{\mathrm{low}},  f_{1\mathrm{yr}} \bigr\}, \qquad
f_{\mathrm{fin}}=\min \bigl\{ f_{\mathrm{high}},  f_{\mathrm{ISCO}} \bigr\}, 
\end{equation}
where 
\begin{equation}
f_{\mathrm{1yr}}=4.149\times 10^{-5} \left[ \left(\frac{\mathcal{M}}{10^6 M_{\odot}}\right)^{-5/8} \left( \frac{T_{\mathrm{obs}}}{1\mathrm{yr}} \right)^{-3/8} \right] \label{f1year}
\end{equation}
is the frequency 1 yr before the coalescence, and $f_{\mathrm{ISCO}}$ is given by 

\begin{equation}
f_{\mathrm{ISCO}}=\frac{1}{6^{3/2}\pi M_{\mathrm{BH}}}.
\end{equation}

For the waveforms, we use the \textit{restricted 2PN waveforms}.
We only take up to 2PN because spin terms are known only to this order.
There are 15 parameters in total:
the chirp mass $\ln\mathcal{M}$, the dimensionless mass parameter $\ln\eta$,
the coalescence time $t_c$, the coalescence phase $\phi_c$,
the distance to the source $\ln D$,
the spin-orbit coupling coefficient $\beta$, 
the spin-spin coupling coefficient $\sigma$,
the inner product of the orbital angular momentum and the total spin angular momentum $\kappa$, 
the precession angle $\alpha_c$ that characterises $\alpha$ at the time of coalescence (the precession angle $\alpha$ is defined in Appendix~\ref{app-prec}), 
the dimensionless asymptotic eccentricity invariant $I_e$, 
$(\theta_{\mathrm{S}},\phi_{\mathrm{S}})$ for the initial direction of the source, 
$(\theta_{\mathrm{J}},\phi_{\mathrm{J}})$ for the initial direction of the total angular momentum,
and finally, $\bar{\omega}$, the inverse of the Brans-Dicke parameter,  or $\beta_g$, the quantity that is proportional to the square inverse of the graviton Compton wavelength $\lambda_g$, depending on which theory we are aiming to constrain.
For constraining the Brans-Dicke parameter, we consider NS/intermediate mass black hole (IMBH) binaries, assuming the difference between NS and BH sensitivities $\mathcal{S}$ to be 0.3.
For constraining the graviton mass, we consider SMBH/BH binaries at 3Gpc.
In this case, we fix the cosmological parameters as follows:
$\Omega_{\kappa}=\Omega_{\mathrm{rad}}=0$, $\Omega_{m}=0.3$, $\Omega_{\Lambda}=0.7$.
We take the Hubble constant to be $H_0=72$ km/s/Mpc. 
There are three main differences from Ref.~\cite{berti}.
 
 (i) We include the spin-spin interaction $\sigma$:
Berti et al.~\cite{berti} reported that when they included both $\sigma$ and the parameter for alternative theories of gravity like $\bar{\omega}$, the ratio (we denote this by $R$) between the smallest and the largest eigen values of the Fisher matrix $\mathbf{\Gamma}$ approached their machine's floating-point precision, and they could not obtain enough accuracy for the inversion of this matrix.
We evade this problem by performing our numerical calculations in quadruple precisions.
Also, we use the trick explained in Appendix~\ref{app-inv} to make sure the numerical inversion is correctly performed.
Basically, we rescale the basis vectors so as to make all the diagonal components of the Fisher matrix equal to 1~\cite{barack}, then take the inverse of this normalised matrix, and finally multiply the factor for the rescaling back to obtain the inverse of our original Fisher matrix.
Even if the ratio $R$ for the original Fisher matrix approaches the machine's floating-point precision, the one for the normalised Fisher matrix stays smaller than the floating-point.
To check that our matrix inversion has been performed correctly, we followed a similar procedure described in Berti \textit{et al}.~\cite{berti}.
We also mention this in Appendix~\ref{app-inv} in more detail.

(ii) We take the eccentricity into account:
The binary system gradually loses energy and angular momentum because of the gravitational radiation which circularises the orbit.
In fact, since the eccentricity decreases following $e\propto f^{-19/18}$~\cite{peters}, usually one does not include the eccentricity into binary parameters assuming that it is 0 \textit{a priori}.
Still, we should take the eccentricity into account for more practical analysis.
%The values that we take for $e_0$ is explained in Appendix~\ref{app1}.

(iii) We also take the spin precession into account:
In this case, the orbital angular momentum $\bm{L}$ oscillates, which introduces the additional oscillations in both the amplitude and the phase of the waveform.
This additional information solves degeneracy in the binary parameters, enhancing the determination accuracy.
The parameter estimation including precession has been estimated by several authors using LISA~{\cite{vecchio,lang} and using the detectors on the ground~\cite{van1,van2,raymond,van3}, in the framework of general relativity.
Our calculation is the first one to include precession in parameter estimation in the context of modified gravity.
We consider the simple precession approximation by assuming that one of the spins of the binary constituents is 0.
References~\cite{vecchio,van1,van2,raymond,van3} estimated the determination errors of the binary parameters under this simple precession approximation.

\subsection{No precession}
\label{subsec-noprec}

\quad In this subsection, we show the results without spin precession effects.
In this case, $\kappa$ and $\alpha_c$ are excluded from our binary parameters.
Following Ref.~\cite{berti}, we perform the numerical integration using the Gauss-Legendre routine GAULEG~\cite{numerical recipes}.
Gauss-Legendre quadrature takes the abscissas at the zeros of the $N$-th Legendre polynomials.
With this method, integrand polynomials up to ($2N-1$)-th order can be calculated exactly.
We take the number of frequency bins to be 600.
The fiducial values of the parameters are $t_c=\phi_c=\beta=\sigma=\bar{\omega}=\beta_g=0$.
%We change the masses of the binary constituents and calculate the eccentricity $e_0$ for each case (see Appendix~\ref{app1}).
%The frequency $f_0$ at which the eccentricity takes the value $e_0$ is chosen as $f_0=0.3$Hz for NS/BH binaries and $f_0=10^{-4}$Hz for SMBH/BH binaries.

In Ref.~\cite{hopman}, Hopman and Alexander performed Monte Carlo simulations in which they followed a star on a relativistic orbit and added small perturbations to simulate energy dissipation and random two-body scattering.
For (1.4+10$^3$)$M_{\odot}$ NS/BH binaries, the probability distribution of eccentricity at $f=2\times 10^{-4}$Hz peaks at $e=0.998$.
From this condition, we further evolve the orbit to yield $e=0.026$ at 1 yr before coalescence.
(See Appendix~\ref{app1} for more details on calculations of the eccentricity evolution.)
For SMBH binaries, Armitage and Natarajan~\cite{armitage} have simulated the interaction between a binary and its surrounding circumbinary gas disk.
For  $M=10^6 M_{\odot}$ SMBH binaries, the eccentricity at 1 yr before coalescence is $e=0.02$ for the equal mass binaries and $e=0.04$ for the unequal mass binaries with mass ratio 0.1.
Recently, Berentzen \textit{et al.}~\cite{berentzen} performed $N$-body simulations of SMBH binaries in rotating galactic nuclei.
They followed the evolution from kiloparsec separations to the final relativistic coalescence, including post-Newtonian corrections up to 2.5.
They found that when SMBH binary reaches the separation of 100 Schwarzschild radii, the typical eccentricity becomes $e=0.04$.
When the total binary mass is $M=10^6 M_{\odot}$, the gravitational wave frequency at $r=100$ Schwarzschild radii is $f=\sqrt{M/(\pi^2 r^3)}=2.25 \times 10^{-5}$Hz.
This frequency is roughly the same as the one at 1 yr before coalescence.
Taking these results into account, we set the fiducial values of $e_0$ as $e_0=0.01$ at $f_0=f_{\mathrm{1yr}}$ for both NS/BH and SMBH binaries.

For the estimations without precession, we assume that the fiducial values of the binary spins are zero.
In these cases, the total angular momentum $\bm{J}$ is equivalent to the orbital angular momentum $\bm{L}$.
Therefore we take the direction of the orbital angular momentum $(\theta_{\mathrm{L}},\phi_{\mathrm{L}})$ as binary parameters instead of $(\theta_{\mathrm{J}},\phi_{\mathrm{J}})$. 
We treat the angles $(\theta_{\mathrm{S}},\phi_{\mathrm{S}})$ and $(\theta_{\mathrm{L}},\phi_{\mathrm{L}})$ in two different ways.
First, we take the average of these angles and calculate the determination accuracy of the binary parameters.
We call this pattern-averaged estimate and we use only 1 detector for the analysis.
The second one is the Monte Carlo simulation.
We randomly distribute $10^4$ binaries ($10^4$ sets of these angles), calculate the determination accuracy for each binary, and take the average at the end.
In this case, we use 2 detectors for the analysis.

\subsubsection{Pattern-averaged estimates }
\label{sec-num-ave}

\quad In the case of pattern-averaged estimates, there are only 9 parameters in total: $\ln \mathcal{M},\ln \eta, t_c,\phi_c,D_L,\beta,\sigma,I_e$ and $\bar{\omega}$ or $\beta_g$.
The waveform is given by Eq.~(\ref{wave-noangle}).
The derivative of this waveform with respect to each parameter is taken analytically as shown in Appendix~\ref{derivative}.

\vspace{3mm}

\begin{table*}
\caption{\label{table-bd-noangle} The results of error estimation in Brans-Dicke theory.
These are calculated with pattern-averaged analysis for NS/BH binaries with BH masses $400 M_{\odot}$, $1000 M_{\odot}$, $5000 M_{\odot}$ and $10000 M_{\odot}$ for $\rho=10$. 
We fix NS masses to $1.4 M_{\odot}$.
For each binary, the first line represents the results shown in~\cite{berti} which does not include $\sigma$ nor $I_e$ into parameters. 
The second line represents the results taking $\sigma$ into account, and the third line shows the ones that include both $\sigma$ and $I_e$.
Here $I_e$ is the value of $e^2$ at $f=0.3$Hz.
We used only one detector for the analyses.}
%\begin{center}
\begin{ruledtabular}
\begin{tabular}{ccc||cccccccc}  %\hline
 & $\sigma$& $I_e$ & $\omega_{\mathrm{BD}}$ & $\Delta \ln\mathcal{M}$(\%) & 
                    $\Delta \ln\eta $ & $\Delta \beta $ &
                    $\Delta t_c(s)$ & $\Delta \phi_c$ & $\Delta \sigma$ & $\Delta I_e (10^{-12})$ \\ \hline\hline 
\multicolumn{3}{c||}{400$M_{\odot}$} & \multicolumn{7}{l}{ } \\
& $\times$ & $\times$ & 39190 & 0.00657 & 0.0250 & 0.0508 & 7.95 & 76.7 & - & - \\ 
& $\bigcirc$ & $\times$ & 24886 & 0.0130 & 0.0819 & 0.202 & 13.8 & 552 & 2.39 & -\\
& $\bigcirc$ & $\bigcirc$ & 4583 & 0.0396 & 0.142 & 0.280 & 16.7 & 552 & 2.49 & 1.09 \\ \hline
\multicolumn{3}{c||}{1000$M_{\odot}$} & \multicolumn{7}{l}{ } \\
& $\times$ & $\times$ & 21257 & 0.00764 & 0.0186 & 0.0557 & 7.99 & 58.4 & - & - \\
& $\bigcirc$ & $\times$ & 8210 & 0.0265 & 0.110 & 0.0692 & 23.5 & 919 & 1.96 & - \\
& $\bigcirc$ & $\bigcirc$ & 1881 & 0.0692 & 0.193 & 0.261 & 23.6 & 1059 & 2.41& 6.34 \\ \hline
\multicolumn{3}{c||}{5000$M_{\odot}$} & \multicolumn{7}{l}{ } \\
& $\times$ & $\times$ &  6486 & 0.0114 & 0.0133 & 0.0550 & 8.79 & 23.4 & - & - \\
& $\bigcirc$ & $\times$ & 1933 & 0.0503 & 0.0936 & 0.221 & 37.9 & 1108 & 0.595 & - \\
& $\bigcirc$ & $\bigcirc$ & 281 & 0.224 & 0.302 & 0.916 & 62.9 & 2438 & 1.30 & 173 \\ \hline
\multicolumn{3}{c||}{10000$M_{\odot}$} & \multicolumn{7}{l}{ } \\
& $\times$ & $\times$ &  3076 & 0.0178 & 0.0161 & 0.0706 & 13.6 & 15.5 & - & - \\
& $\bigcirc$ & $\times$ & 862 & 0.0827 & 0.114 & 0.350 & 82.9 & 1763 & 0.474 & - \\
& $\bigcirc$ & $\bigcirc$ & 113 & 0.412 & 0.418 & 1.51 & 160 & 4454 & 1.11& 797 \\ %\hline
\end{tabular}
\end{ruledtabular}
%\end{center}
\end{table*}

\noindent
\textbf{Brans-Dicke theory.-}

\quad Here, we consider the parameter estimation in the context of Brans-Dicke theory. 
We think of four NS/BH binaries of SNR=10. 
When performing Fisher matrix analysis, this SNR might be too small~\cite{vallisneri} and we should use the binaries with much larger SNR.
In that case, since the constraint on $\omega_{\mathrm{BD}}$ is proportional to SNR, all we have to do is to appropriately scale the results obtained for SNR=10. 
We fix $m_{\mathrm{NS}}=1.4$M$_{\odot}$ and take $m_{\mathrm{BH}}=400$, 1000, 5000 and 10$^4$ M$_{\odot}$. 
For each binary, we show three results in Table~\ref{table-bd-noangle}: the first line represents the determination accuracies of binary parameters without taking $\sigma$ and $I_e$ into parameters (the same estimates by Berti \textit{et al.}~\cite{berti}), the second line shows the results including $\sigma$ but not $I_e$ into parameters, and the third line shows the ones including both $\sigma$ and $I_e$ into parameters.

\begin{figure}[thbp]
  \includegraphics[scale=.4,clip]{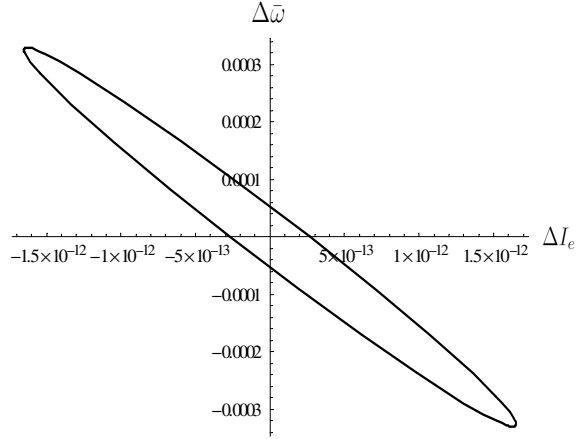} 
 \caption{\label{e-bd} The curve showing 68$\%$ confidence level on the $\Delta e^2$-$\Delta \bar{\omega}$ plane for a $(1.4+400)M_{\odot}$ NS/BH binary with $\rho=10$, such that the fiducial values lie at the centre of the ellipse.}
\end{figure}

From this table, one can see that adding parameters reduces the determination accuracy.
This is because the parameters are strongly correlated and adding parameters dilutes the binary information in the detected gravitational waves.
Including both $\sigma$ and $I_e$ into parameters increases the determination errors by roughly 1 order of magnitude.
In particular, including $I_e$ increases the errors on $\omega_{\mathrm{BD}}$ more than just including $\sigma$.
The reason for this is as follows.
In the phase $\Psi(f)$, the term containing $\omega_{\mathrm{BD}}$ is proportional to $f^{-2/3}$, whilst the ones containing $\sigma$ and $I_e$ have the frequency dependence $f^{4/3}$ and $f^{-19/9}$, respectively.
The terms containing $\omega_{\mathrm{BD}}$ and $I_e$ both have negative power-law indices and the dependences on the frequency are similar.
Therefore $\omega_{\mathrm{BD}}$ has stronger correlation with $I_e$ than $\sigma$.

Comparing the constraints from the binaries having different BH mass, one can see that the constraint becomes more stringent as the BH mass decreases.
This is because the orbital velocity of the binaries become slower, which makes the dipole contribution relatively greater.   

In Fig.~\ref{e-bd}, we plot the curve of constant probability on the $\Delta I_e$-$\Delta \bar{\omega}$ plane for a $(1.4+400)M_{\odot}$ NS/BH binary with $\rho=10$.
The fiducial values lie at the centre of the presented ellipse at 68$\%$ confidence level.
Since $I_e$ is always positive, it can be understood from this figure that we can decrease the upper bound on $\Delta \bar{\omega}$ by imposing the prior distribution $I_e>0$.
In fact, the upper limit for $\Delta\bar{\omega}$ is obtained at $\Delta I_e=0$ in this case.
This means that when we include the above prior information, the constraint on $\omega_{\mathrm{BD}}$ including eccentricity should be similar to the case without taking eccentricity into account.
Therefore we conclude that the constraint on $\omega_{\mathrm{BD}}$ is not so much affected by taking eccentricity into account.

\vspace{3mm}

\begin{table*}
\caption{\label{table-massive-noangle}The results of error estimation in massive graviton theories. These are calculated with pattern-averaged analysis for SMBH/BH binaries with masses $(10^7+10^7)M_{\odot}$, $(10^7+10^6)M_{\odot}$, $(10^6+10^6)M_{\odot}$ and $(10^6+10^5)M_{\odot}$ at 3Gpc. As in the Brans-Dicke case, the first line of each binary represents the estimation without taking $\sigma$ nor $I_e$ into parameters.
These results do not exactly match with the ones shown in~\cite{berti} because we also considered prior information. The meaning of second and third lines are the same as in Table~\ref{table-bd-noangle}.}
\begin{ruledtabular}
\begin{tabular}{ccc||cccccccc}  %\hline
 & $\sigma$& $I_e$ & $\lambda_g$($10^{20}$cm) & $\Delta \ln\mathcal{M}$(\%) & 
                    $\Delta \ln\eta $ & $\Delta \beta $ &
                    $\Delta t_c(s)$ & $\Delta \phi_c$ & $\Delta \sigma$ & $\Delta I_e (10^{-10})$ \\ \hline\hline 
\multicolumn{3}{c||}{$(10^7+10^7)M_{\odot}$} & \multicolumn{7}{l}{ } \\
& $\times$ & $\times$ & 22.77 & 0.0669 & 0.467 & 2.93 & 75.7 & 1.06 & - & - \\ 
& $\bigcirc$ & $\times$ & 11.33 & 0.0687 & 0.960 & 7.10 & 77.8 & 1.09 & 1.73 & -\\
& $\bigcirc$ & $\bigcirc$ & 11.29 & 0.246 & 1.10 & 7.56 & 133 & 2.23 & 1.89 & 13.0 \\ \hline
\multicolumn{3}{c||}{$(10^7+10^6)M_{\odot}$} & \multicolumn{7}{l}{ } \\
& $\times$ & $\times$ & 9.629 & 0.0493 & 0.253 & 1.49 & 82.3 & 2.15 & - & - \\
& $\bigcirc$ & $\times$ & 4.061 & 0.0495 & 0.839 & 6.26 & 82.8 & 2.16 & 1.89 & - \\
& $\bigcirc$ & $\bigcirc$ & 4.052 & 0.202 & 0.927 & 6.54 & 159 & 5.01 & 1.94& 14.3 \\ \hline
\multicolumn{3}{c||}{$(10^6+10^6)M_{\odot}$} & \multicolumn{7}{l}{ } \\
& $\times$ & $\times$ &  12.41 & 0.00869 & 0.122 & 0.787 &  3.01& 0.316 & - & - \\
& $\bigcirc$ & $\times$ & 3.582 & 0.00871 & 0.926 & 6.97 & 3.01 & 0.316 & 1.68 & - \\
& $\bigcirc$ & $\bigcirc$ & 3.582 & 0.0288 & 0.933 & 7.00 & 4.75 & 0.589 & 1.69 & 1.14 \\ \hline
\multicolumn{3}{c||}{$(10^6+10^5)M_{\odot}$} & \multicolumn{7}{l}{ } \\
& $\times$ & $\times$ &  6.019 & 0.00586 & 0.0551 & 0.337 & 2.45 & 0.521 & - & - \\
& $\bigcirc$ & $\times$ & 1.286 & 0.00586 & 0.823 & 6.20 & 2.45 & 0.521 & 1.88 & - \\
& $\bigcirc$ & $\bigcirc$ & 1.285 & 0.0200 & 0.826 & 6.21 & 3.96 & 0.989 & 1.88 & 1.50 \\ %\hline
\end{tabular}
\end{ruledtabular}
\end{table*}

\noindent
\textbf{Massive Graviton theories.-}

\quad Next, we consider the parameter estimation in the context of massive graviton theories. 
We also consider four BH/BH binaries in this case: $(10^7+10^7)M_{\odot}, (10^7+10^6)M_{\odot}, (10^6+10^6)M_{\odot}$ and $(10^6+10^5)M_{\odot}$.
As in the Brans-Dicke case, we perform three types of analyses whose results are shown in Table~\ref{table-massive-noangle}.
Berti \textit{et al.}~\cite{berti} reported that the effect of adding prior information on the maximum spin is negligible.
We confirmed that including prior information changes the results just about a few percent.
Berti \text{et al.} have also claimed that when they include both $\sigma$ and $\bar{\omega}$ or $\beta_g$ into parameters, they cannot take the inverse of the Fisher matrix properly.
This is true but, if we include the prior information, the matrix inverse can be performed successfully.
Hence, we decided to include the prior information.
This is the reason why the first column of each binaries in Table~\ref{table-massive-noangle} slightly differs from the results shown in Ref.~\cite{berti}.

\begin{figure}[thbp]
  \includegraphics[scale=.4,clip]{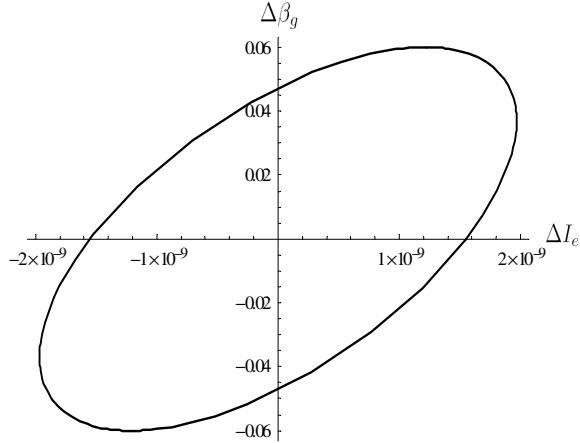} 
 \caption{\label{e-massive} The curve showing 68$\%$ confidence level on the $\Delta e^2$-$\Delta \beta_g$ plane for a $(10^7+10^7)M_{\odot}$ SMBH/BH binary at 3Gpc. The fiducial values lie at the centre of the ellipse as in Fig.~\ref{e-bd}.}
\end{figure}

From Table~\ref{table-massive-noangle}, it can be seen that in this case, including both $\sigma$ and $I_e$ into parameters reduces the accuracy only by a factor of a few.
This indicates that $\beta_g$ is only weakly correlated with other parameters.
In the massive graviton case, including $\sigma$ affects the constraint on $\lambda_g$ more than including $I_e$.
The reason is the same as in the case of Brans-Dicke.
Since both terms containing $\lambda_g$ and $\sigma$ in the phase $\Psi(f)$ have frequency dependences with positive power-law indices, their correlation is stronger than the one between $\lambda_g$ and $I_e$.

Figure~\ref{e-massive} shows the curve of constant probability on the $\Delta I_e$-$\Delta \beta_g$ plane for a $(10^7+10^7)M_{\odot}$ SMBH/BH binary at 3Gpc.
The fiducial values lie at the centre of the ellipse at 68$\%$ confidence level.
Compared to Fig.~\ref{e-bd} of the Brans-Dicke theory, it can be seen that the ellipse is slightly tilted in the other way.
This indicates that the constraint on $\lambda_g$ would not change even if we impose the prior distribution $\Delta I_e >0$.
However, this ellipse is tilted only slightly, which suggests that $\beta_g$ and $I_e$ are correlated only weakly, as we have stated above.
Therefore, the constraint on $\lambda_g$ is affected only weakly by the inclusion of $I_e$ into parameters.

When the graviton has a finite mass, its phase velocity is expressed as~\cite{will1998}

\begin{equation}
v_{\mathrm{ph}}^2=\left( 1-\frac{1}{f^2\lambda_g^2}\right)^{-1}.
\end{equation}  
Therefore, at a lower frequency the difference between $v_{\mathrm{ph}}$ and the speed of light ($c=1$) is larger.
This is why the heavier binaries put a more stringent constraint on $\lambda_g$.

\subsubsection{Estimates without pattern-averaging}
\label{sec-num-mc}

\quad In the case of estimates without pattern-averaging, there are 13 parameters in total: the 9 parameters appeared in the last subsection and 4 angles representing the source direction and the inclination of the orbit.
We randomly generate $10^4$ sets of the four angles so that $\cos\theta_{\mathrm{S}}$ and $\cos\theta_{\mathrm{L}}$ are uniformly distributed in the range $[-1,1]$ and $\phi_{\mathrm{S}}$ and $\phi_{\mathrm{L}}$ in the range $[0,2\pi]$.
We use the RAN1 routine~\cite{numerical recipes} for generating random numbers and calculate the parameter estimation errors for each binary.
We take the average at the end. 

We take the parameter derivatives of the part $e^{i\Psi(f)}$ in the waveform analytically as in the pattern-averaged case, whilst we take the derivatives of the rest numerically.
The angular resolution $\Delta\Omega_{\mathrm{S}}$ is defined as

\begin{equation}
\Delta\Omega_S\equiv 2\pi \lvert \sin\bar{\theta}_{\mathrm{S}} \rvert \sqrt{\Sigma_{\bar{\theta}_{\mathrm{S}}\bar{\theta}_{\mathrm{S}}}
                              \Sigma_{\bar{\phi}_{\mathrm{S}}\bar{\phi}_{\mathrm{S}}}-\Sigma^2_{\bar{\theta}_{\mathrm{S}}\bar{\phi}_{\mathrm{S}}}}.
\end{equation}
We consider the following two cases: including not $I_e$ but $\sigma$ into binary parameters, and including both $\sigma$ and $I_e$ into parameters.

Following Ref.~\cite{berti}, we display the results in histograms.
For each binary parameter, we obtain $10^4$ error values.
We group them into $N_{\mathrm{bins}}$ bins as follows;
if an error value $X$ satisfies 

\begin{widetext}
\begin{equation}
\left[ \ln(X_{\min})+\frac{(j-1)[\ln(X_{\max})-\ln(X_{\min})]}{N_{\mathrm{bins}}} \right] < \ln(X) 
                               <\left[ \ln(X_{\min})+\frac{j[\ln(X_{\max})-\ln(X_{\min})]}{N_{\mathrm{bins}}} \right],
\end{equation}
\end{widetext}
then we define that this belongs to the $j$-th bin.
We divide the number of binaries of each bin by the total number of binaries 10000 to get a "probability distribution."
We show this probability against the determination error of each parameter in a histogram below.

\vspace{3mm}

\begin{table*}
\caption{\label{table-bd-noprec} The results of error estimation in Brans-Dicke theory for $(1.4+1000)M_{\odot}$ NS/BH binaries without pattern averaging. We used two detectors for the analyses and we fixed $\rho=\sqrt{200}$ ($\rho=10$ for each detector). We performed the following Monte Carlo simulations. We distribute $10^4$ binaries, calculate the error of each parameter for each binary and take the average. The first half of the table shows the results without taking precessional effect, and the second half represents the ones including precession. The first line of each part shows the results without taking $I_e$ into parameters, whilst in the second line this is taken into account. $\sigma$ is included in the parameter for all the cases. }
\begin{ruledtabular}
\begin{tabular}{c||ccccccc}  %\hline
 Cases & $\omega_{\mathrm{BD}}$ &  $\Delta \ln\mathcal{M}(\%)$  & 
                    $\Delta\ln \eta$ & $\Delta \beta $ &
                    $\Delta \ln D_L$ & $\Delta \Omega_S(10^{-3}\mathrm{str})$ & $\Delta \sigma$  \\ \hline\hline
% &  & ($\%$) & & & & ($10^{-3}$str) & \\ \hline\hline 
%\multicolumn{6}{l}{(1.4+400)$M_{\odot}$} \\
\textbf{No precession} & \multicolumn{6}{l}{ } \\
Excluding $I_e$ & 4844 &  0.0396 & 0.143 & 0.108 & 2.53 & 0.406 & 2.18   \\ 
Including $I_e$ & 1058 & 0.106 & 0.241 & 0.481 & 1.34 &1.05 & 2.45  \\ \hline
\textbf{Including precession} & \multicolumn{6}{l}{ } \\
Excluding $I_e$ & 6944 & 0.0291 & 0.107 & 0.144 & 0.0809 & 0.341 & 1.66  \\
Including $I_e$ & 3523 & 0.0432 & 0.130 & 0.161 & 0.0851 & 0.589 & 1.86  \\ %\hline
\end{tabular}
\end{ruledtabular}
\end{table*}

\noindent
\textbf{Brans-Dicke theory.-}

\begin{figure}[htbp]
  \includegraphics[scale=.5,clip]{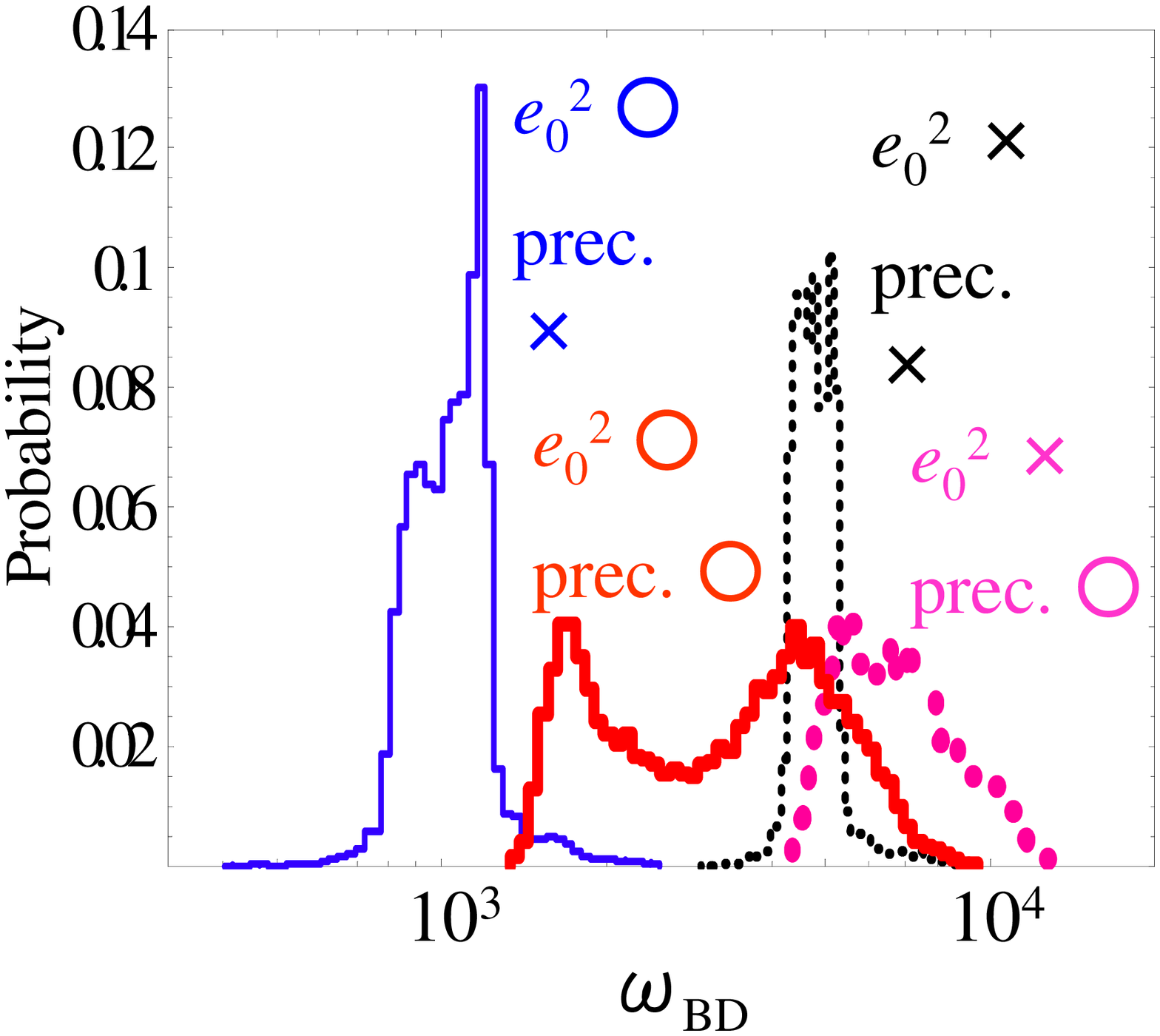} 
 \caption{\label{bd-new} The histograms showing the probability distribution of the lower bound of $\omega_{\mathrm{BD}}$ obtained from the Monte Carlo simulations with $(1.4+1000)M_{\odot}$ NS/BH binaries with $\rho=\sqrt{200}$ in Brans-Dicke theory. The (black) dotted thin one represents the estimate without precession and $I_e$ is not taken into account. The (blue) solid thin one shows the one without precession but $I_e$ is taken into parameters. The (purple) dotted thick  one includes precession but does not include $I_e$. The (red) solid thick curve shows the one including both precession and $I_e$. Here the prior information $I_e>0$ is not used to estimate the lower bound  of $\omega_{\mathrm{BD}}$}
\end{figure}

\begin{figure*}[htbp]
 \includegraphics[scale=.7,clip]{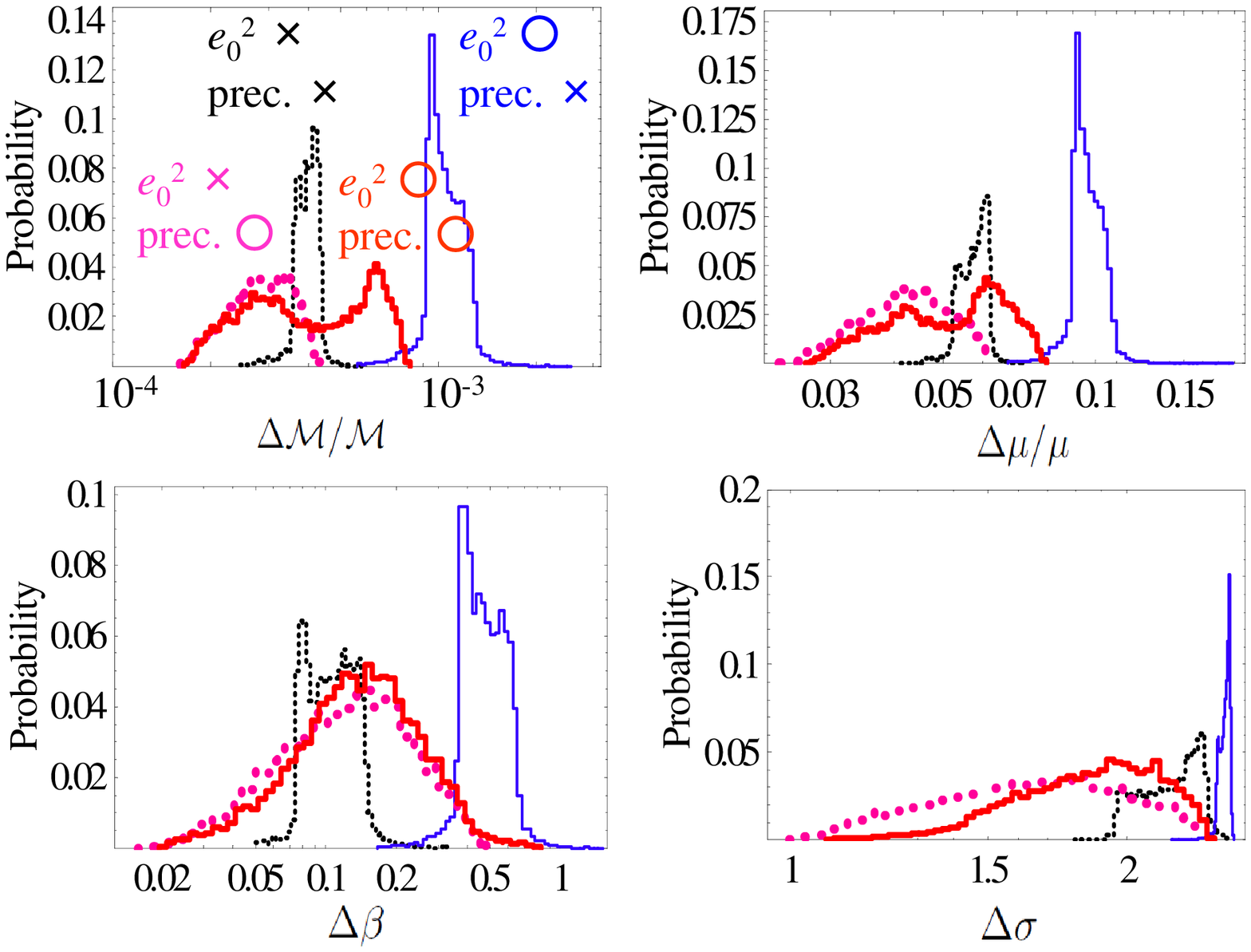} 
\caption{\label{bd-mc}
The histograms showing the probability distribution of the error estimates of the chirp mass $\Delta \mathcal{M}/\mathcal{M}$, the reduced mass $\Delta\mu/\mu$, the spin-orbit coupling $\Delta\beta$, and the spin-spin coupling $\Delta\sigma$ in Brans-Dicke theory. These are obtained from the Monte Carlo simulations with $(1.4+1000)M_{\odot}$ NS/BH binaries with $\rho=\sqrt{200}$. The meaning of different types of curves is the same as in Fig.~\ref{bd-new}.}
\end{figure*}

\begin{figure*}[htbp]
 \includegraphics[scale=.7,clip]{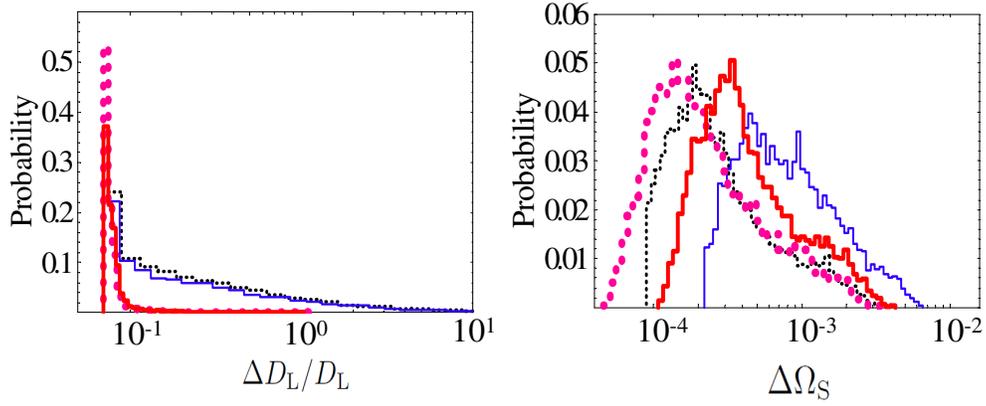} 
\caption{\label{bd-omega}
The histograms showing the probability distribution of the error estimates of the luminosity distance $\Delta D_L/D_L$ and the angular resolution $\Delta\Omega_{S}$ in Brans-Dicke theory. These are obtained from the Monte Carlo simulations with $(1.4+1000)M_{\odot}$ NS/BH binaries with $\rho=\sqrt{200}$. The meaning of different types of curves is the same as in Fig.~\ref{bd-new}.}
\end{figure*}

\begin{figure*}[htbp]
  \includegraphics[scale=.4,clip]{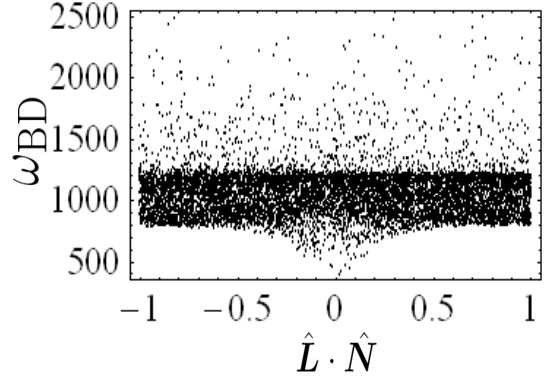} 
 \caption{\label{bd-ln}
$10^4$ plots of the lower bounds of $\omega_{\mathrm{BD}}$ against $\hat{\bm{L}}\cdot\hat{\bm{N}}$. This result is obtained from the Monte Carlo simulations with $(1.4+1000)M_{\odot}$ NS/BH binaries with $\rho=\sqrt{200}$ in Brans-Dicke theory. $I_e$ is taken into account though the precessional effect is not included.}
\end{figure*}

\begin{figure*}[htbp]
  \includegraphics[scale=.8,clip]{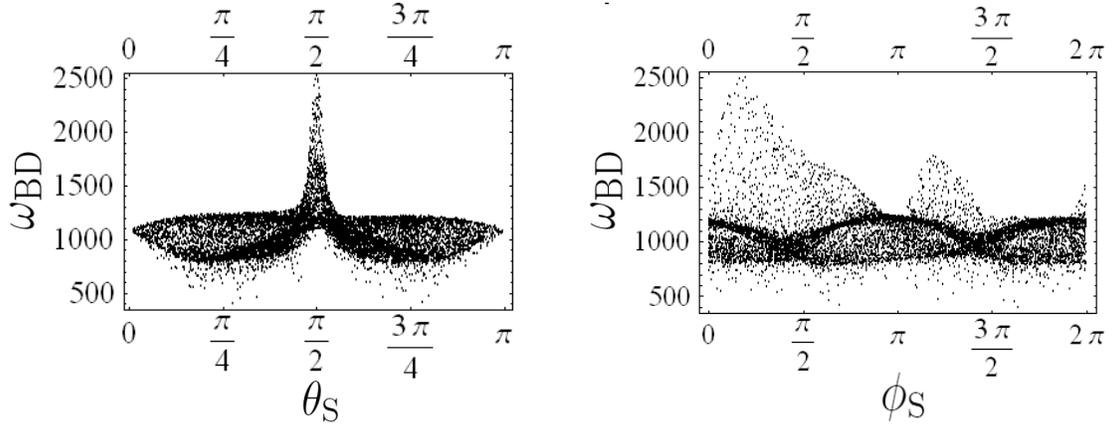} 
 \caption{\label{bd-angles}
$10^4$ plots of the lower bounds of $\omega_{\mathrm{BD}}$ against $\bar{\theta}_{\mathrm{S}}$ and $\bar{\phi}_{\mathrm{S}}$, as in Fig.~\ref{bd-ln}.}
\end{figure*}

\begin{figure*}[htbb]
  \includegraphics[scale=.4,clip]{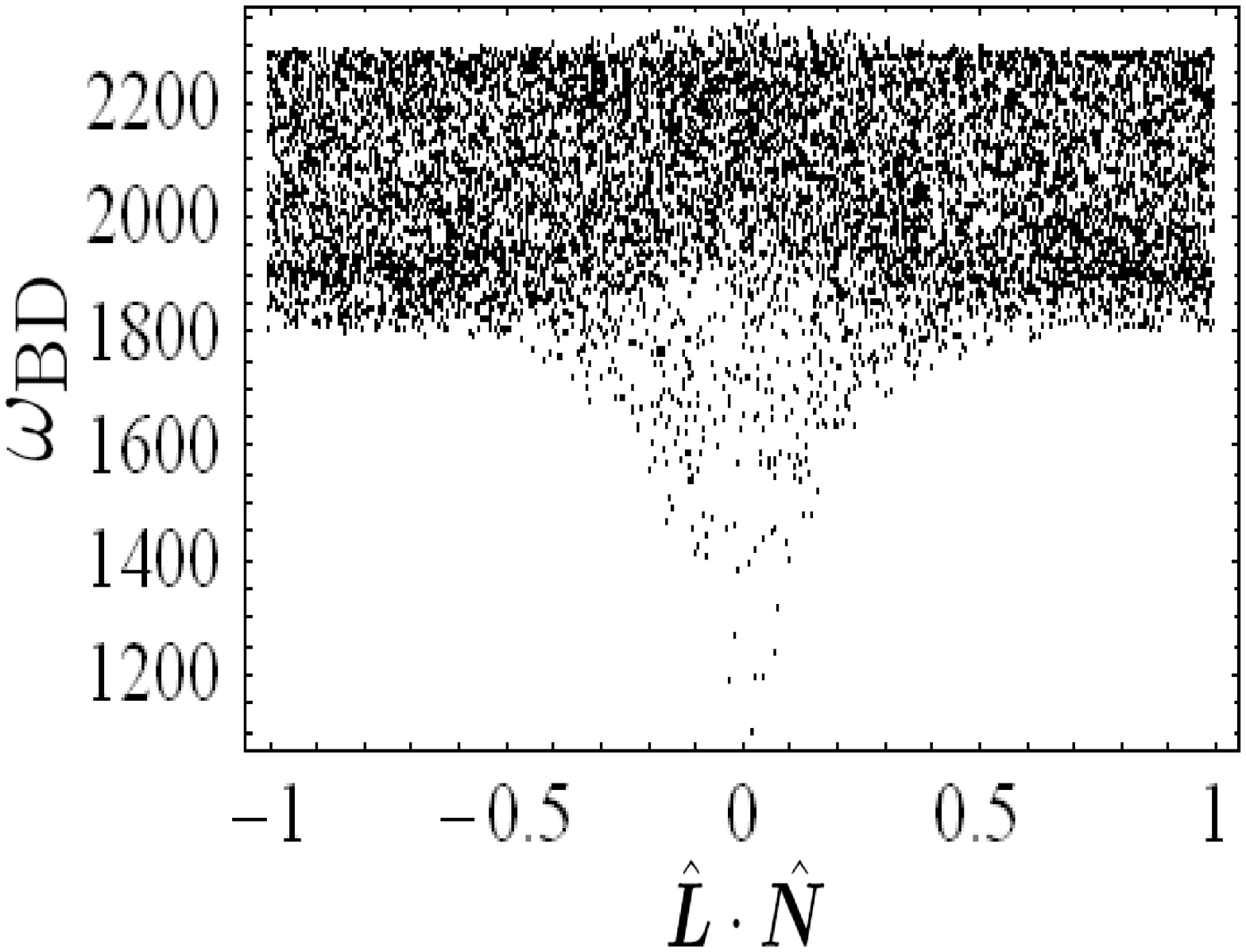} 
 \caption{\label{bd-ln-nodp}
The same plot as Fig.~\ref{bd-ln} except we do not include the Doppler phase $\varphi_D(t)$ in this case.}
\end{figure*}

%\clearpage
\begin{figure*}[htbp]
  \includegraphics[scale=.8,clip]{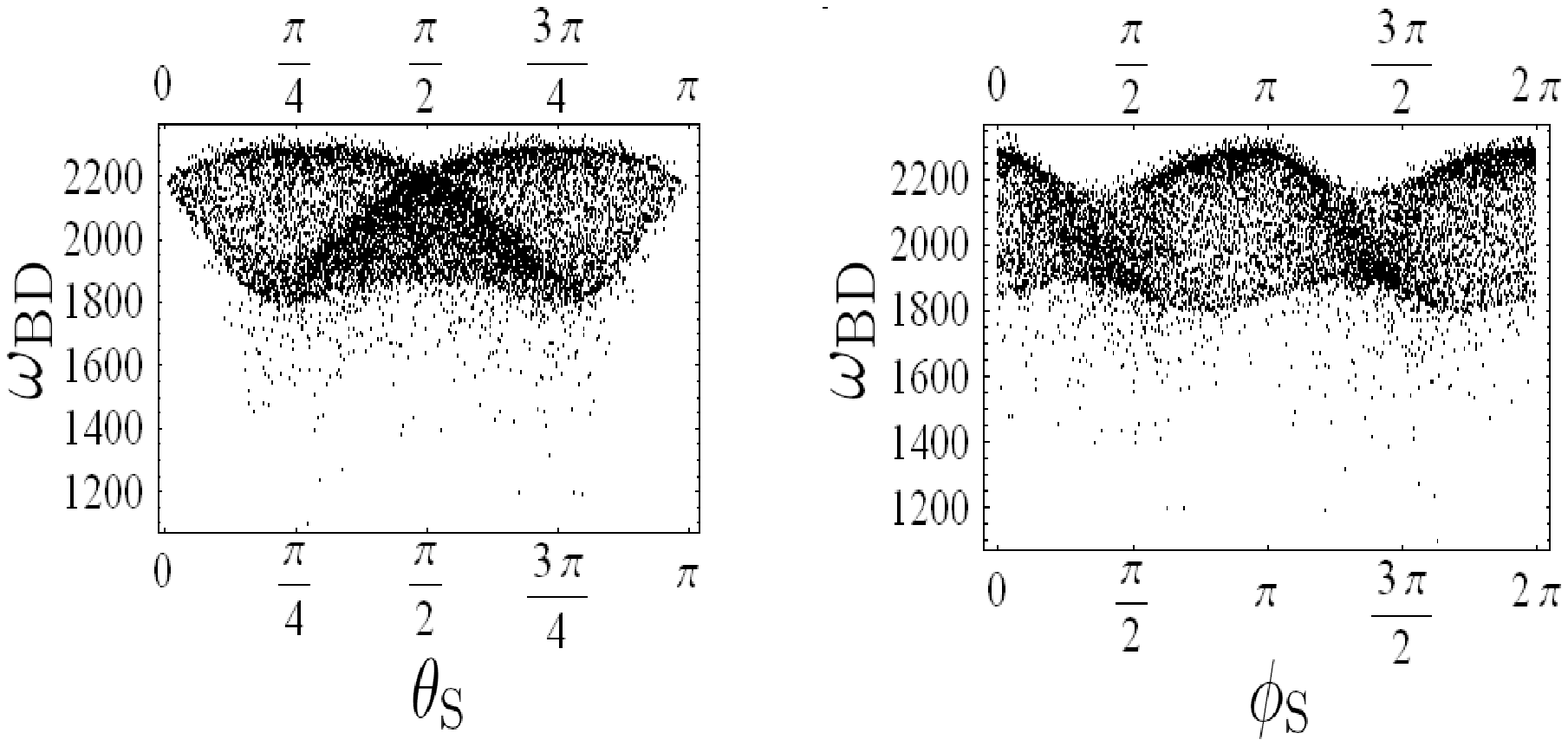} 
 \caption{\label{bd-angles-nodp}
The same plot as Fig.~\ref{bd-angles} except we do not include the Doppler phase $\varphi_D(t)$ in this case.}
\end{figure*}

We fix the masses of the binary constituents as $m_{\mathrm{NS}}=1.4 M_{\odot}$ and $m_{\mathrm{BH}}=10^4 M_{\odot}$.
We consider binaries of $\rho=\sqrt{200}$ (roughly speaking this corresponds to $\rho=10$ for a single detector).
We show the results in the upper half of Table~\ref{table-bd-noprec}.
The first row represents the error estimation without including $I_e$ as a fitting parameter and the second row shows the one including $I_e$.
The corresponding histograms are shown in Figs.~\ref{bd-new},~\ref{bd-mc} and~\ref{bd-omega}.  
The (black) dotted thin line represents the results without including $I_e$ as a binary parameter and the (blue) solid thin line shows the one including $I_e$. 
The (purple) dotted thick  and the (red) solid thick lines are the ones including precession, which are to be explained later.
For the moment, let us focus on the (black) dotted thin and the (blue) solid thin lines only.
Figure~\ref{bd-new} is the histogram of the lower bound of $\omega_{\mathrm{BD}}$.
Including $I_e$ increases the error by a factor of 4, as expected from the results of pattern-averaged estimation.
Figure~\ref{bd-mc} shows the histograms of $\Delta \mathcal{M}/\mathcal{M}$, $\Delta\eta/\eta$, $\Delta\beta$, and $\Delta \sigma$.
These accuracies also get worse by several factors when $I_e$ are taken into account.
Figure~\ref{bd-omega} shows the histograms of $\Delta D_L/D_L$ and $\Delta\Omega_S$.
%The error estimations of $\Delta D_L/D_L$ with and without including precession are shown separately.
The (black) dotted thin and (blue) solid thin lines in the histogram of $\Delta D_L/D_L$ are almost the same.
As $D_L$ appears only in the amplitude of the waveform, this parameter is uncorrelated with the phase parameters such as $I_e$.
The histograms of $\Delta D_L/D_L$ have tails on the lower accuracy side.
These tails are due to the binaries with $\hat{\bm{L}}\cdot\hat{\bm{N}}=\pm 1$ because then the distance and the direction of $\hat{\bm{L}}$ degenerate.
The estimation errors for $\Delta D_L/D_L$ go up to $10^4$ for our calculation but we only show them up to $10$ in Fig.~\ref{bd-omega}.
Again, the accuracy of $\Delta\Omega_S$ goes down by several factors when we include $I_e$.

In Figs.~\ref{bd-new} and~\ref{bd-mc}, there are also some tails in the histograms.
The ones with worse accuracy are due to the binaries with $\hat{\bm{L}}\cdot\hat{\bm{N}}=0$.
Figure~\ref{bd-ln} shows $10^4$ plots of the constraints on $\omega_{\mathrm{BD}}$ against $\hat{\bm{L}}\cdot\hat{\bm{N}}$, taking eccentricity into account (these plots correspond to the (blue) solid thin histogram in Fig.~\ref{bd-new}).
It can be seen that the constraints get worse when the binaries become $\hat{\bm{L}}\cdot\hat{\bm{N}}=0$.
The constraints on $\omega_{\mathrm{BD}}$ against $\theta_{\mathrm{S}}$ and $\phi_{\mathrm{S}}$ are shown in Fig.~\ref{bd-angles}. 
It is clear that binaries with some special source directions $(\theta_{\mathrm{S}},\phi_{\mathrm{S}})$ have the parameter estimation accuracy enhanced. 
These tails do not depend on the directions of $\bm{L}$.
We found that they are due to the motion of the detectors (the Doppler phase $\varphi_{D}(t)$).
The Eq.~(\ref{doppler-phase}) for $\varphi_{D}(t)$ can be recast into the form~\cite{kocsis}

\begin{equation}
\varphi_{D}(t)=2\pi \frac{f(t)}{f_c} \sin \bar{\theta}_{\mathrm{S}} \cos[\bar{\phi}(t)-\bar{\phi}_{\mathrm{S}}], 
\end{equation}
with the critical frequency $f_c$ defined by 

\begin{equation}
f_c\equiv \frac{c}{R} =2.00\mathrm{mHz},
\end{equation}
for $R=1$AU. 
This shows that the effect of $\varphi_{D}(t)$ is important for frequencies higher than 2mHz.
Since the relevant frequency range for a $(1.4+1000)M_{\odot}$ binary is $3.66\times 10^{-2}-1.00\mathrm{Hz}$, $\varphi_{D}(t)$ has some remarkable contribution. 
In Figs.~\ref{bd-ln-nodp} and~\ref{bd-angles-nodp}, we show the same plots but without including the Doppler phase. 
It is clear that the anomalous peaks have disappeared in this case.
The comparison of the constraint on $\omega_{\mathrm{BD}}$ between these cases suggest that the degeneracies among some parameters become strong when we include $\varphi_D$.
However, these degeneracies are solved in some special cases which correspond to the peaks in Fig.~\ref{bd-angles}.
For example, the reason why $\bar{\theta}_{\mathrm{S}}=\frac{\pi}{2}$ is special can be understood as follows.
Since $\varphi_D(t)$ is proportional to $\sin\bar{\theta}_{\mathrm{S}}$, the derivative of $\varphi_D$ with respect to $\bar{\theta}_{\mathrm{S}}$ is proportional to $\cos\bar{\theta}_{\mathrm{S}}$.
Therefore when $\bar{\theta}_{\mathrm{S}}=\frac{\pi}{2}$, this term vanishes and there would be no degeneracy between $\omega_{\mathrm{BD}}$ and $\bar{\theta}_{\mathrm{S}}$ that is caused by the Doppler phase.

%\newpage

\vspace{3mm}

%\begin{table*}
%\caption{\label{table-massive-noprec} The results of the Monte Carlo simulations in massive graviton theories for $(10^7+10^6)M_{\odot}$ BH/BH binaries at 3Gpc without pattern averaging. We used two detectors for the analyses. As in the Brans-Dicke case, we distribute $10^4$ binaries, calculate the error of each parameter for each binary and take the average. The meaning of each row is the same as in Table~\ref{table-bd-noprec}.}
%\begin{ruledtabular}
%\begin{tabular}{c||cccccccc}  %\hline
% Cases & $\lambda_g(10^{21}\mathrm{cm})$ & SNR & $\Delta \ln\mathcal{M}(\%)$  & 
%                    $\Delta\ln \eta$ & $\Delta \beta $ &
 %                   $\Delta \ln D_L$ & $\Delta \Omega_S(10^{-4}\mathrm{str})$ & $\Delta \sigma$  \\ \hline\hline
% & ($10^{21}$cm) & & ($\%$) & & & & ($10^{-4}$str) & \\ \hline\hline 
%\textbf{No precession} & \multicolumn{7}{l}{ } \\
%Excluding $I_e$ & 1.2507 & 1540 & 0.0507 & 0.147 & 0.834 & 0.0235 & 1.05 &  0.402 \\ 
%Including $I_e$ & 1.1906 & 1540 & 0.199 & 0.343 & 1.56 & 0.0238 & 1.07 &  0.720 \\ \hline
%\textbf{Including precession} & \multicolumn{7}{l}{ } \\
%Excluding $I_e$ & 4.8645 & 1589 & 0.00834 & 0.00675 & 0.0118 & 0.00188 &  0.354 & 0.0508 \\
%Including $I_e$ & 3.0954 & 1594 & 0.0262 & 0.00697 & 0.0119 & 0.00191 & 0.366 &  0.0808 \\ %\hline
%\end{tabular}
%\end{ruledtabular}
%\end{table*}

\begin{table*}
\caption{\label{table-massive-noprec} The results of the Monte Carlo simulations in massive graviton theories for $(10^7+10^6)M_{\odot}$ BH/BH binaries at 3Gpc without pattern averaging. We used two detectors for the analyses. As in the Brans-Dicke case, we distribute $10^4$ binaries, calculate the error of each parameter for each binary and take the average. The meaning of each row is the same as in Table~\ref{table-bd-noprec}.}
\begin{ruledtabular}
\begin{tabular}{c||cccccccc}  %\hline
 Cases & $\lambda_g(10^{21}\mathrm{cm})$ & SNR & $\Delta \ln\mathcal{M}(\%)$  & 
                    $\Delta\ln \eta$ & $\Delta \beta $ &
                    $\Delta \ln D_L$ & $\Delta \Omega_S(10^{-4}\mathrm{str})$ & $\Delta \sigma$  \\ \hline\hline
% & ($10^{21}$cm) & & ($\%$) & & & & ($10^{-4}$str) & \\ \hline\hline 
\textbf{No precession} & \multicolumn{7}{l}{ } \\
Excluding $I_e$ & 0.40598 & 1540 & 0.0507 & 0.841 & 6.27 & 0.0230 & 0.957 &  1.89 \\ 
Including $I_e$ & 0.40507 & 1540 & 0.191 & 0.927 & 6.54 & 0.0233 & 0.972 &  1.94 \\ \hline
\textbf{Including precession} & \multicolumn{7}{l}{ } \\
Excluding $I_e$ & 4.8540 & 1596 & 0.00838 & 0.00675 & 0.0117 & 0.00189 &  0.366 & 0.0508 \\
Including $I_e$ & 3.0570 & 1586 & 0.0269 & 0.00708 & 0.0120 & 0.00192 & 0.364 &  0.0825 \\ %\hline
\end{tabular}
\end{ruledtabular}
\end{table*}

\begin{figure}[htbp]
  \includegraphics[scale=.45,clip]{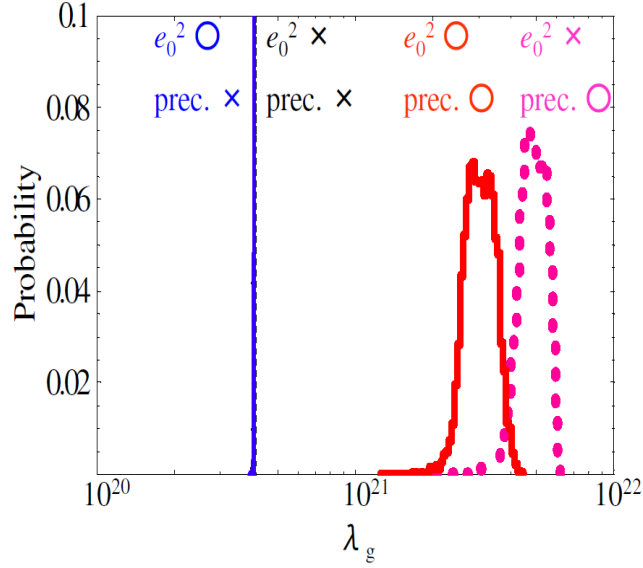} 
 \caption{\label{massive-lambda}
The histograms showing the probability distribution of the lower bound of $\lambda_g$ obtained from the Monte Carlo simulations with $(10^7+10^6)M_{\odot}$ BH/BH binaries at 3Gpc in massive graviton theories. The (black) dotted thin one represents the estimate without precession and $I_e$ is not taken into account. The (blue) solid thick  one shows the one without precession but $I_e$ is taken into parameters. The (purple) dotted thin one includes precession but does not take $I_e$ into account. The (red) solid thick one shows the one including precession and $I_e$ is also taken into account.}
\end{figure}

\begin{figure*}[htbp]
  \includegraphics[scale=.65,clip]{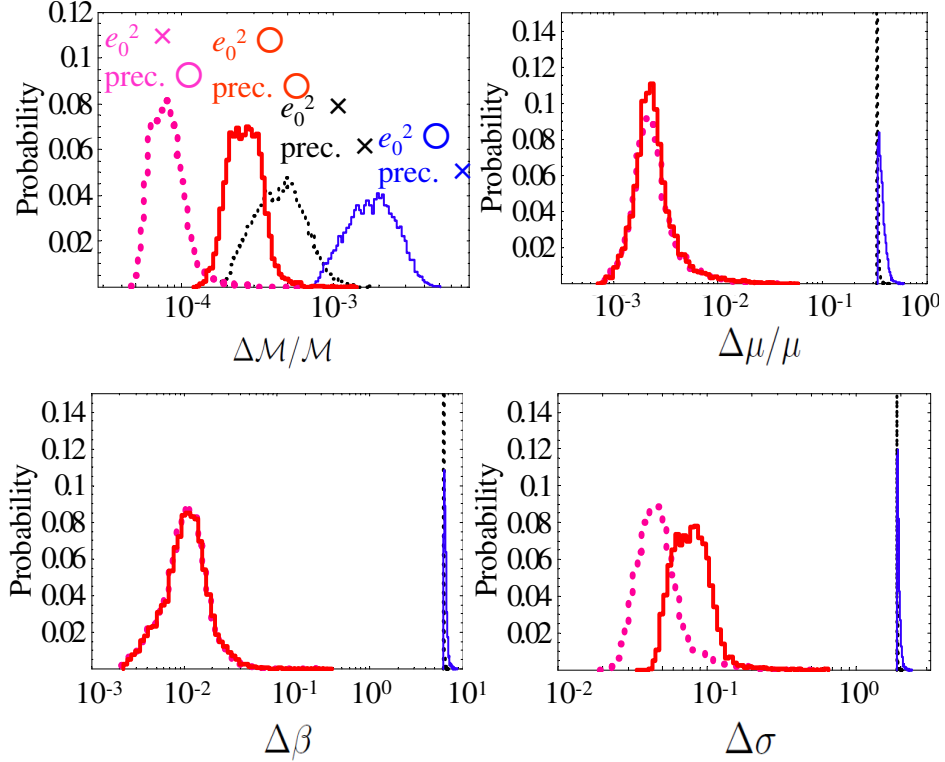} 
\caption{\label{massive-mc}
The histograms showing the probability distribution of the error estimates of the chirp mass $\Delta \mathcal{M}/\mathcal{M}$, the reduced mass $\Delta\mu/\mu$, the spin-orbit coupling $\Delta\beta$ and the spin-spin coupling $\Delta\sigma$ in massive graviton theories. These are obtained from the Monte Carlo simulations with $(10^7+10^6)M_{\odot}$ BH/BH binaries at 3Gpc. The meaning of different types of curves is the same as in Fig.~\ref{massive-lambda}.}
\end{figure*}

\begin{figure*}[htbp]
 \includegraphics[scale=.65,clip]{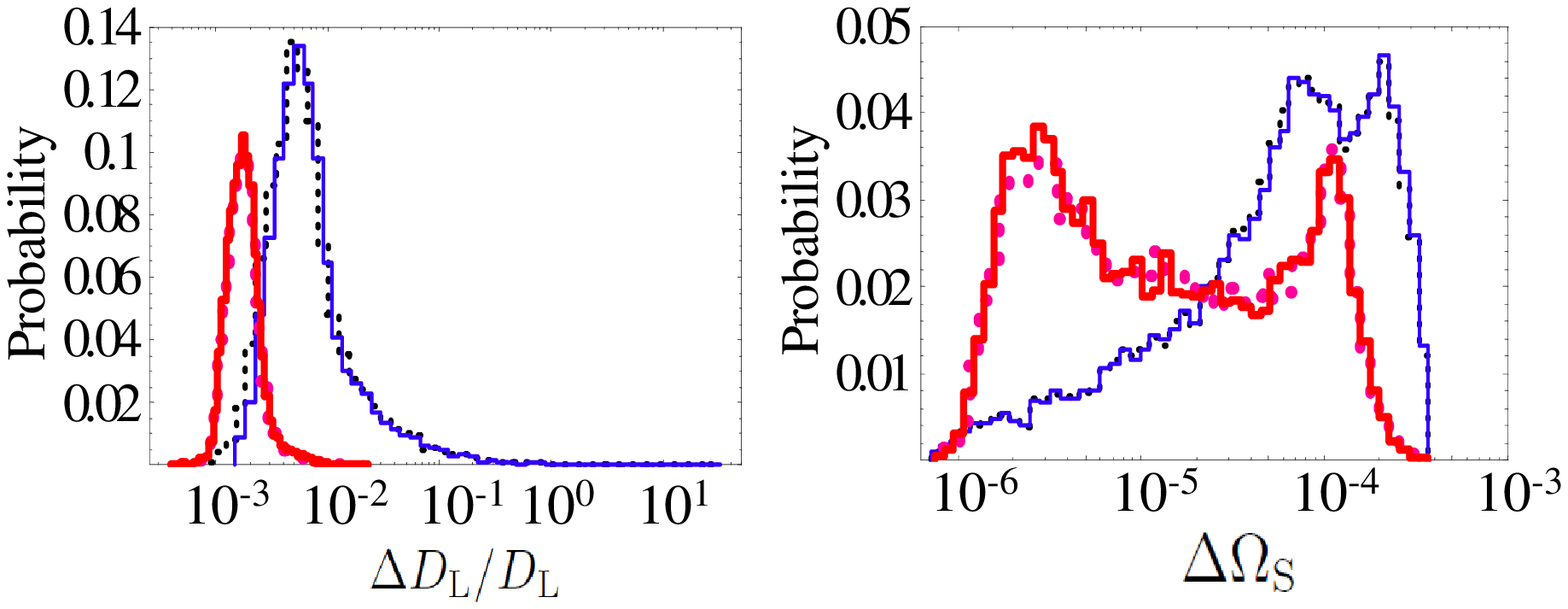} 
\caption{\label{massive-omega}
The histograms showing the probability distribution of the error estimates of the luminosity distance $\Delta D_L/D_L$ and the angular resolution $\Delta\Omega_{S}$ in massive graviton theories. These are obtained from the Monte Carlo simulations with $(10^7+10^6)M_{\odot}$ BH/BH binaries at 3Gpc. The meaning of different types of curves is the same as in Fig.~\ref{massive-lambda}. The accuracy of the luminosity distance with and without precession are shown separately.}
\end{figure*}

\noindent
\textbf{Massive Graviton theories.-}

We fix the masses of the binary constituents to $10^7 M_{\odot}$ and $10^6 M_{\odot}$, and the distance $D_L=$3Gpc.
The results are shown in the upper half of Table~\ref{table-massive-noprec}.
As in the Brans-Dicke case, the first row represents the error estimation without including $I_e$ into binary parameters and the second row shows the one including $I_e$.
As in the Brans-Dicke case, the corresponding histograms are shown in Fig.~\ref{massive-lambda} for the lower bound of $\lambda_g$, in Fig.~\ref{massive-mc} for the error estimations of masses and spins, and in Fig.~\ref{massive-omega} for the ones of the distances and the source directions. 
Again, the (black) dotted thin line represents the results without including $I_e$ as a binary parameter and the (blue) solid thin line shows the ones including $I_e$. 
In the massive graviton case, the parameter estimation is only weakly dependent on the inclusion of parameter $I_e$.

There are 5 parameters in the phase $\Psi (f)$ shown in Eq.~(\ref{Psi-noangle}); $\mathcal{M},\eta,\beta,\sigma$, and $\beta_g$. 
On the other hand, there are only 4 PN terms in total (the leading quadrupole term, 1PN, 1.5PN, and 2PN).
This means that these 5 parameters degenerate when we do not include precession. 
Then, one of the parameters is totally unconstrained without prior distribution. 
The dispersion that the histograms for $\lambda_g$, $\mu$, $\beta$ and $\sigma$ have in Figs.~\ref{massive-lambda} and~\ref{massive-mc} is mainly determined by the level of uncertainty in the determination of the spin parameter $\sigma$. 
When we do not take into account the precession, the level of uncertainty of $\sigma$ is totally determined by the prior distribution, and hence it is identical for all binaries. 
Therefore each histogram for $\lambda_g$, $\mu$, $\beta$ and $\sigma$ has a very sharp peak when we do not include the precession effect.

%As seen from Figs.~\ref{massive-lambda} and~\ref{massive-mc}, in contrast to the Brans-Dicke case, we notice that there are no tails on the better accuracy side.
%This is because the frequency range for a $(10^7+10^6)M_{\odot}$ binary, $2.36\times 10^{-5}-4.00\times 10^{-4}\mathrm{Hz}$, is lower than $f_c=2$mHz, and the doppler phase has negligible contribution to the results.

\subsection{Including Precession}
\label{sec-num-prec}

\quad In this subsection, we show the results when we include the spin precession.
There are 15 parameters in total.
We perform the Monte Carlo simulations as in the previous subsection.
Here, we set the dimensionless spin parameter of the lighter body of binaries to 0 and that of the heavier body to 0.5.  
We randomly distribute the fiducial values of $\kappa=\hat{\bm{L}}\cdot\hat{\bm{S}}$ in the range [-1,1] and choose $\alpha_c$ randomly in the range $[0,2\pi]$.
We use $L=\mu\sqrt{M r}$ for the calculation of orbital angular momentum, where the separation is given as $r=M^{1/3}/(\pi f)^{2/3}$.
(This is derived from $v^2=\sqrt{M/r}=(\pi M f)^{2/3}$.)
When we include the spin precession, the waveform gets some additional oscillations.
This worsens the precision of the polynomial approximation to the waveform. 
For this reason, we decide not to use the Gauss-Legendre quadrature for the numerical integrations.
Instead, we cut the integrand into $10^4$ pieces, equally binned in terms of $f^{-1}$.  

\vspace{3mm}

\noindent
\textbf{Brans-Dicke theory.-}

We fix the masses of the binary constituents to $m_{\mathrm{NS}}=1.4 M_{\odot}$ and $m_{\mathrm{BH}}=10^4 M_{\odot}$ and SNR $\rho$ to $\sqrt{200}$.
We neglect the spins for the neutron stars, which is supported from observations~\cite{blanchet}. 
For the dimensionless spin parameters of the black holes, we adopt $\chi=0.5$.
The results are shown in the lower half of Table~\ref{table-bd-noprec}.
The first row in this table represents the results without taking eccentricity $I_e$ into parameters and the second row shows the results including $I_e$.
The corresponding histograms are again shown in Figs.~\ref{bd-new},~\ref{bd-mc} and~\ref{bd-omega}. 
The (purple) dotted thick  line represents the estimation without taking $I_e$ into parameters and the (red) solid thick one is the estimation with $I_e$.
Figure~ \ref{bd-new} shows that the constraints on $\omega_{\mathrm{BD}}$ increase by 20$\%$ when the precession effect is taken into account.
In~\cite{berti}, Berti \textit{et al.} found that by detecting $(1.4+1000)M_{\odot}$ NS/BH binary gravitational waves of $\rho=\sqrt{200}$ with LISA, one can put a constraint $\omega_{\mathrm{BD}}>10799$ on average.
Comparing this result with ours, it is understood that including the effects of $\sigma$, $I_e$ and precession, in total, leads to weaken the constraint by an order of magnitude.
Therefore these effects cannot be neglected.
However, as we stated before, imposing the prior distribution $\Delta I_e >0$ would reduce the upper limit of $\Delta\bar{\omega}$ to the case without including $I_e$ into parameters.
This means that this prior information strengthens the constraint on $\omega_{\mathrm{BD}}$ by a factor of 6. 
In this case, the constraint on $\omega_{\mathrm{BD}}$ becomes $\omega_{\mathrm{BD}}>6944$, which is just 1.6 times lower than the one obtained in~\cite{berti}.
Figure~\ref{bd-new} shows that when eccentricity is taken into account, the effect of precession is larger than in the case without including it.
This is because the degeneracy between parameters are disentangled by precession.
In Fig.~\ref{bd-mc}, one can see that the lower accuracy tails disappear when we include precession, as there is no binary with $\hat{\bm{L}}\cdot\hat{\bm{N}}$ fixed to $0$ due to the precession of $\hat{\bm{L}}$.
Also in Fig.~\ref{bd-omega}, the tails on the histograms disappear when we include precession for the same reason.

We have checked that when we take the limit $\kappa =1$, our Monte Carlo simulation reduces to the one without including precession.
However, we found that when we set $\kappa=0.999$, the constraint on $\omega_{\mathrm{BD}}$ differs from the one with $\kappa=1$ by several factors.
This unphysical fact shows that we cannot estimate the errors properly with Fisher analysis in some regions.
Above $\kappa = 0.999$ (and below $\kappa = -0.999$), the estimation error of precession angle $\Delta \alpha$ (not $\Delta \alpha_c$) exceeds 2$\pi$.
This means that the effect of precession is too weak to be determined and the linear analysis of Fisher matrix is invalid.
Therefore we have to say that our calculation cannot be trusted for these regions.  
When we exclude these regions and recalculate the average for the constraint on $\omega_{\mathrm{BD}}$, it becomes $\omega_{\mathrm{BD}}>3252$ when we include eccentricity.
Comparing this with the corresponding result in Table~\ref{table-bd-noprec}, $\omega_{\mathrm{BD}}>3251$, we can say that the contribution of these untrustable regions are not so important.

In Fig.~\ref{bd-mc}, the results of $\Delta\beta$ may look a bit strange.
When $I_e$ is not included into parameters, the determination accuracy gets worse when we include spin precession.
This is due to the difference of the fiducial values that we are taking.
When we do not consider the effect of spin precession, the fiducial value of the dimensionless BH spin parameter $\chi_{BH}$ is set to 0.
On the other hand, when we include spin precession, it is set to 0.5.
If we set this BH spin parameter to 0.5 and perform the Monte Carlo simulation without including precession, the histogram gets shifted.
This is shown in Fig.~\ref{bd-spin-0.5}.
The (black) dotted thin one represents the estimate without spin precession and the fiducial value for $\chi_{\mathrm{BH}}$ is set to 0, whilst the (black) solid one represents the one with $\chi_{\mathrm{BH}}=0.5$ and $\kappa =1$. 
The (purple) dotted thick  one represents the one including spin precession and the fiducial value for $\chi_{\mathrm{BH}}$ is set to 0.5.
Comparing the (black) solid histogram and the (purple) dotted thick  one, it can be seen that the inclusion of spin precession enhances the determination accuracy of $\beta$ just like for the other parameters. 
For the case of $\chi_{\mathrm{BH}}=0.5$, we obtained the constraint on $\omega_{\mathrm{BD}}$ as $\omega_{\mathrm{BD}}>4854$ on average. 
Comparing this to the one obtained when we set $\chi_{\mathrm{BH}}=0$, $\omega_{\mathrm{BD}}>4862$, it can be seen that the offset of spin parameter affects the constraint on $\omega_{\mathrm{BD}}$ only a little.

\begin{figure}[tbp]
 \includegraphics[scale=.35,clip]{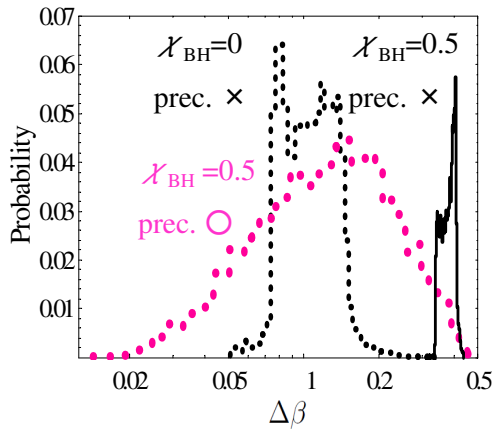} 
\caption{\label{bd-spin-0.5}
The histograms showing the probability distribution of the error estimates of the spin-orbit coupling $\Delta\beta$ in Brans-Dicke theory. These are obtained from the Monte Carlo simulations with $(1.4+1000)M_{\odot}$ NS/BH binaries with $\rho=\sqrt{200}$.  Eccentricity is not included into the binary parameters. The (black) dotted thin one represents the estimate without spin precession and the fiducial value for $\chi_{\mathrm{BH}}$ is set to 0, whilst the (black) solid one represents the one with $\chi_{\mathrm{BH}}=0.5$. The (purple) dotted thick  one represents the one including spin precession and the fiducial value for $\chi_{\mathrm{BH}}$ is set to 0.5.}
\end{figure}

\vspace{3mm}

\noindent
\textbf{Massive Graviton theories.-}

We fix the masses of the binary constituents to $10^7 M_{\odot}$ and $10^6 M_{\odot}$.
We assume the spins of the $10^6 M_{\odot}$ black holes to be 0 for simplicity.
As in the case of Brans-Dicke theory, we take the spins of heavier black holes to be $\chi=0.5$. 
Under this assumption, we can apply the simple precession approximation. 
The results are shown in the second half of Table~\ref{table-massive-noprec}.
The third row in this table represents the results without taking eccentricity $I_e$ into parameters and the fourth row shows the ones including $I_e$.
The corresponding histograms are shown in Figs.~\ref{massive-lambda},~\ref{massive-mc} and~\ref{massive-omega}. 
The (purple) dotted thick  line represents the estimation without taking $I_e$ into parameters and the (red) solid thick one is the estimation with $I_e$.
In Fig.~\ref{massive-lambda}, it is shown that the lower bounds on $\lambda_g$ increase by one order of magnitude when we include precessional effect.
Compared to the result of~\cite{berti}, which states that the detection of $(10^6+10^6)M_{\odot}$ BH/BH inspiral gravitational waves with LISA leads to the constraint $\lambda_g>1.33\times 10^{21}$cm on average, our result shows that the inclusion of $\sigma$, $I_e$ and the precessional effect in total enhances the constraint by a factor of 2.3.
In this case, since the parameter estimation only depends weakly on $\sigma$ and $I_e$, the important point is to include the precession which makes the constraint stronger.
Figure~\ref{massive-mc} shows that the inclusion of precession considerably enhances the accuracy of $\Delta \mu/\mu$ and $\Delta\beta$ by 2 or 3 orders of magnitude.
In Fig.~\ref{massive-omega}, one can see again that the tails on $\Delta D_L/D_L$ disappear due to the change in $\hat{\bm{L}}$.  
Histograms in Figs.~\ref{massive-lambda} and~\ref{massive-mc} have tails on the lower accuracy side.
These correspond to binaries with $\kappa \approx 1$, where the effects of precession almost reduce to zero.

The reason why the effect of spin precession is larger for the massive graviton case than the Brans-Dicke one is because the additional information other than the phase $\Psi(f)$ is crucial for the parameter estimation in the massive graviton case.
The phase of restricted 2PN waveform $\Psi(f)$ contains five binary parameters but it has only four PN terms.
This makes these parameters degenerate.
This degeneracy is solved to different levels for different binaries when we include precession, which makes the dispersion in the histograms broader.
On the other hand, in the Brans-Dicke case, there are 5 PN terms in the 2PN phase $\Psi(f)$.
Therefore, the degeneracies between parameters are not so strong and the effect of precession is relatively weak.

%The amount of precessional effect is characterised by the precession angle $\alpha$, defined in Eq.~(\ref{alpha-omega}).
%$\alpha\propto f^{-1}$ for the case $L \gg S$ whilst $\alpha\propto f^{-2/3}$ for the case $L \ll S$~\cite{apostolatos}.
%Therefore, the effect of precession dominates in the low frequency band.
%For the analysis of Brans-Dicke theory, the frequency range of $(1.4+1000)M_{\odot}$ is $(f_{\mathrm{in}},f_{\mathrm{fin}})=(3.66\times 10^{-2}\mathrm{Hz},1.00\mathrm{Hz})$ and about 80$\%$ of SNR is accumulated during lower frequency range between $f_{\mathrm{in}}$ and 0.1Hz.
%This means that LISA is already sensitive in lower frequency region without including precession.
%Therefore inclusion of it does not enhance the estimation accuracy much.
%On the other hand, for the analysis of massive graviton theories, the frequency range of $(10^7+10^6)M_{\odot}$ is $(f_{\mathrm{in}},f_{\mathrm{fin}})=(2.36\times 10^{-5}\mathrm{Hz},4.00\times 10^{-4}\mathrm{Hz})$.
%In this case, more than half of SNR comes from higher frequency range between $3\times 10^{-4}$ and $f_{\mathrm{fin}}$.
%This indicates that LISA is not sensitive in lower frequency region and the accuracy of parameter estimation increases drastically when we include spin precession.

\section{CONCLUSIONS}
\label{sec-conclusion}

\quad In this paper, we extended the previous analysis by Berti \textit{et al.}~\cite{berti} to see how the inclusion of the spin-spin coupling $\sigma$, the eccentricity $e_0$ and the precessional effect affects the binary parameter estimation by means of LISA in the context of alternative theories of gravity such as the Brans-Dicke and massive graviton theories.
For the Brans-Dicke case, we assumed that we detect NS/BH inspiral gravitational waves with SNR=$\sqrt{200}$ because the binaries composed of different types of compact objects enhance the gravitational dipole radiation. 
On the other hand, we thought of BH/BH inspiral gravitational waves at 3Gpc for the massive graviton case because the remarkable difference between $v_{\mathrm{ph}}$ and $c$ appears in lower frequency GWs.

First, we performed the analysis using the pattern-averaged waveform. 
For the error estimation in Brans-Dicke theory, the inclusion of both $\sigma$ and $I_e$ into parameters reduces the determination accuracy by an order of magnitude.
In particular, including $I_e$ affects the estimation more than including just $\sigma$.
However, if we impose the prior information $\Delta I_e >0$, the constraint becomes stronger and would be as good as the one without including $I_e$ into parameters.
For the analysis in the massive graviton theories, the inclusion of these parameters only changes the results by a factor of a few.
In this case, the inclusion of $\sigma$ affects more than the inclusion of $I_e$.
Also in the massive gravity case, imposing the prior distribution $\Delta I_e$ would not affect the constraint on $\lambda_g$.

Next, we performed the Monte Carlo simulations including precession. 
We set the dimensionless spin parameter $\chi$ of the lighter body of binaries to 0 and that of the heavier body to 0.5, which corresponds to setting the Kerr parameter to the half of mass in the black hole case.
We found that in Brans-Dicke case, the results are not so much affected by taking precession into account.
For a NS/BH binary of $(1.4+1000)M_{\odot}$ with SNR=$\sqrt{200}$, estimation with taking $\sigma$ and precession into account can lead to a constraint $\omega_{\mathrm{BD}}>6944$ on average ($I_e$ is not important for the constraint on $\omega_{\mathrm{BD}}$ because of the prior information for $I_e$). 
However, when we consider the event rate, the detectability of $(1.4+1000)M_{\odot}$ binaries of SNR=$\sqrt{200}$ is very low.
Therefore only a lucky detection can constrain $\omega_{\mathrm{BD}}$.
For example, referring to Brown \textit{et al.}~\cite{brown}, the event rate for EMRI is given by $0.7\times(300M_{\odot}/M)\times 10^{-10}$Mpc$^{-3}$yr$^{-1}$.
The distance of $(1.4+1000)M_{\odot}$ binaries of SNR=$\sqrt{200}$ approximately corresponds to 40Mpc for LISA. 
Therefore the event rate is $7.6\times 10^{-7}$yr$^{-1}$. 
However, for DECIGO and BBO, the event rate of NS/stellar mass BH binaries with SNR=$\sqrt{200}$ is about $10^4$/yr (see~\cite{shibata} and references therein).
Therefore, these binaries are thought to be the definite sources for them.
By using these detectors, we will obtain stronger constraint~\cite{kent}, because (i) the number of gravitational cycles

\begin{equation}
N_{\mathrm{GW}}\equiv\int^{f_{\mathrm{fin}}}_{f_{\mathrm{in}}}df \frac{f}{\dot{f}},
\end{equation}
is larger, (ii) the velocity at 1 yr before coalescence is slower and (iii) the width of effective frequency range of observation is larger.
We can further improve our constraints by using statistical opportunities of large event rates.
We show these results in a separate paper~\cite{kent}. 
%If we consider the binaries whose event rate is 1/yr for DECIGO and BBO, the sources are expected to have a large signal-to-noise ratio and we can put a few orders of magnitude stronger constraint than the solar system experiment.

In the case of massive graviton theory, inclusion of precession has more remarkable effect.
The constraint on $\lambda_g$ now becomes an order of magnitude stronger.
For a BH/BH binary of $(10^7+10^6)M_{\odot}$ at 3Gpc, estimation with taking $\sigma$, $I_e$ and precession into account can constrain the graviton Compton wavelength as $\lambda_g>3.06\times10^{21}$cm on average.
This is 2.3 times stronger than the result obtained in Ref.~\cite{berti}.  
%In the massive graviton case, the constraint on $\lambda_g$ obtained by DECIGO and BBO is weaker than the one obtained using LISA as larger mass binaries put stronger constraint~\cite{kent}.

%From these results, we conclude that the 
Our results are consistent with previous results, although our error estimates might be underestimates since our analysis does not include systematic errors~\cite{cutler-vallisneri}. 

In Ref.~\cite{yunes-ecc}, authors have derived binary waveforms in the frequency domain including higher order eccentricity terms up to $e_0^8$. 
We included these terms and calculated the constraints on both $\omega_{\mathrm{BD}}$ and $\lambda_g$ with pattern-averaged analyses.
We found that with our choice of fiducial eccentricities $e_0=0.01$ at 1 yr before coalescence, the inclusion of higher order effects do not affect the constraints.
These higher order terms affect the constraints on $\omega_{\mathrm{BD}}$ only when $e_0$ is larger than 0.1.
It seems that this rather high eccentricity is not realised at 1 yr before coalescence from the discussion in Sec.~\ref{subsec-noprec} and Appendix~\ref{app1}.
The effect of these higher order terms is much weaker for the massive gravity case as the correlation between eccentricity and $\lambda_g$ is weaker compared to the Brans-Dicke case.  

In our calculation, we neglected the spin of one of the binaries.
To make our analysis more general, we need to take the spins of both binary stars into account.
In that case, because the simple precession approximation is no longer valid anymore, we need to solve the precession equations as Stavridis and Will did~\cite{stavridis}. 
We can also improve our analysis by introducing higher harmonics to the waveform.
Recently, Arun and Will~\cite{arun} estimated the possible constraint on $\lambda_g$ using the higher harmonics waveform, assuming that they detect the SMBH/BH inspiral GWs with LISA, advanced LIGO and ET.
Since they did not take spins into account, we should try to include both precession and higher harmonics into analysis together, and perform the Monte Carlo simulations in alternative theories of gravity such as the Brans-Dicke and the massive graviton theories.

\begin{acknowledgments}
\quad We thank Naoki Seto and Takashi Nakamura for useful discussions and numerical code corrections, and Bernard Schutz for valuable comments.
This work is in part supported by the Grant-in-Aid for Scientific Research No. 19540285 and No. 21244033.
This work is also supported in part by the Grant-in-Aid for the Global COE Program ``The Next Generation of Physics, Spun from Universality and Emergence'' from the Ministry of Education, Culture, Sports, Science and Technology (MEXT) of Japan.
\end{acknowledgments}

\appendix

\section{DETECTED SIGNALS}
\label{app-det-signal}

The interferometer having three arms corresponds to having two individual detectors.
Therefore, it is possible to measure both polarisations with one detector.
We first focus on the detector I which consists of the arms 1 and 2.
This detector measures 

\begin{equation}
h_{\mathrm{I}}(t) \equiv \frac{\delta L_1(t)-\delta L_2(t)}{L} 
        = \frac{\sqrt{3}}{2} \left( \frac{1}{2}h_{xx}-\frac{1}{2}h_{yy} \right), \label{output}   
\end{equation}
where $\delta L_1(t)$ and $\delta L_2(t)$ are the differences in length in arms 1 and 2 as caused by passing gravitational waves.
$L$ is the length of the arms where gravitational waves are not present.
The factor $\sqrt{3}/2$ comes from the 60$^\circ$ opening angle of adjacent arms.
Here, we introduce two principal axes for the wave;
$\hat{\bm{p}}=\hat{\bm{N}}\times\hat{\bm{L}}/\lvert \hat{\bm{N}}\times\hat{\bm{L}} \rvert$ and 
$\hat{\bm{q}}=\hat{\bm{p}}\times\hat{\bm{N}}$, where $\hat{\bm{L}}$ is the unit vector parallel to the orbital angular momentum and $\hat{\bm{N}}$ is the unit vector pointing toward the centre of mass of the binary.
Then, the two polarisations become exactly $\pi/2$ out of phase and the waveform becomes

\begin{equation}
h_{ab}(t)=A_{+} H^+_{ab} \cos\phi(t) + A_{\times} H^{\times}_{ab} \sin\phi(t), 
\end{equation}
where $H^+_{ab}$ and $H^{\times}_{ab}$ are the polarisation basis tensors, defined as

\begin{equation}
H^{+}_{ab}=p_a p_b-q_a q_b, \qquad H^{\times}_{ab}=p_a q_b+q_a p_b. \label{Hpol}
\end{equation}
From Eq.~(\ref{output})-Eq.~(\ref{Hpol}), the detector output $h_{\mathrm{I}}(t)$ becomes

\begin{widetext}
\begin{equation}
h_{\mathrm{I}}(t)=\frac{\sqrt{3}}{2} A_{+} F_{\mathrm{I}}^+(\theta_{\mathrm{S}},\phi_{\mathrm{S}},\psi_{\mathrm{S}}) \cos\phi(t)
          +\frac{\sqrt{3}}{2} A_{\times} F_{\mathrm{I}}^{\times}(\theta_{\mathrm{S}},\phi_{\mathrm{S}},\psi_{\mathrm{S}}) \sin\phi(t). \label{output2}
\end{equation}
%\end{widetext}
$F_{\mathrm{I}}^+(\theta_{\mathrm{S}},\phi_{\mathrm{S}},\psi_{\mathrm{S}})$ and $F_{\mathrm{I}}^{\times}(\theta_{\mathrm{S}},\phi_{\mathrm{S}},\psi_{\mathrm{S}})$ are the detector beam-pattern coefficients and when the detector is an interferometer, they are given by

%\begin{widetext}
\begin{eqnarray}
F_{\mathrm{I}}^{+}(\theta_{\mathrm{S}},\phi_{\mathrm{S}},\psi_{\mathrm{S}}) 
                &=&\frac{1}{2}(1+\cos^2 \theta_{\mathrm{S}}) \cos(2\phi_{\mathrm{S}}) \cos (2\psi_{\mathrm{S}})
                  -\cos(\theta_{\mathrm{S}}) \sin(2\phi_{\mathrm{S}}) \sin(2\psi_{\mathrm{S}}),  \\
F_{\mathrm{I}}^{\times}(\theta_{\mathrm{S}},\phi_{\mathrm{S}},\psi_{\mathrm{S}})
                &=&\frac{1}{2}(1+\cos^2 \theta_{\mathrm{S}}) \cos(2\phi_{\mathrm{S}}) \sin (2\psi_{\mathrm{S}})
                  +\cos(\theta_{\mathrm{S}}) \sin(2\phi_{\mathrm{S}}) \cos(2\psi_{\mathrm{S}}). \label{beam-pattern}
\end{eqnarray} 
\end{widetext}
$(\theta_{\mathrm{S}},\phi_{\mathrm{S}})$ represents the direction of the source in the detector frame and $\psi_{\mathrm{S}}$ is the polarisation angle defined as 

\begin{equation}
\tan\psi_{\mathrm{S}}=\frac{\hat{\bm{q}}\cdot\hat{\bm{z}}}{\hat{\bm{p}}\cdot\hat{\bm{z}}}
                                =\frac{\hat{\bm{L}}\cdot\hat{\bm{z}}-(\hat{\bm{L}}\cdot\hat{\bm{N}})(\hat{\bm{z}}\cdot\hat{\bm{N}})}
                                  {\hat{\bm{N}}\cdot(\hat{\bm{L}}\times\hat{\bm{z}})}. \label{tanpsi}
\end{equation}

Also, we can think of another detector consisting of arms 2 and 3. We call this detector II' and the signal of this detector can be written as $h_{\mathrm{II}'}=(\delta L_2(t)-\delta L_3(t))/L$.
However, since $h_{\mathrm{I}}$ and $h_{II'}$ have some correlations, they are not independent detectors.
We combine detectors I and II' to construct detector II which is uncorrelated with detector I.
The signal of detector II is 

\begin{equation}
h_{\mathrm{II}}(t) \equiv \frac{1}{\sqrt{3}}[h_{\mathrm{I}}(t)+2h_{II'}(t)]=\frac{\sqrt{3}}{2}\left[ \frac{1}{2}(h_{xy}+h_{yx}) \right].
\end{equation}
This detector II corresponds to an interferometer that is rotated by 45$^\circ$ with respect to detector I.
Thus the beam-pattern coefficients for the detector II are

\begin{eqnarray}
F_{\mathrm{II}}^{+}(\theta_{\mathrm{S}},\phi_{\mathrm{S}},\psi_{\mathrm{S}})&=&F_{\mathrm{I}}^{+}(\theta_{\mathrm{S}},\phi_{\mathrm{S}}-\pi/4,\psi_{\mathrm{S}}), \\
F_{\mathrm{II}}^{\times}(\theta_{\mathrm{S}},\phi_{\mathrm{S}},\psi_{\mathrm{S}})&=&F_{\mathrm{I}}^{\times}(\theta_{\mathrm{S}},\phi_{\mathrm{S}}-\pi/4,\psi_{\mathrm{S}}).
\end{eqnarray}
Reexpressing the waveforms measured by each detector in terms of an amplitude and phase, they become Eq.~(\ref{waveform1}).

When we perform parameter estimation, we take the direction of the source $(\bar{\theta}_{\mathrm{S}},\bar{\phi}_{\mathrm{S}})$ and the direction of the orbital angular momentum $(\bar{\theta}_{\mathrm{L}},\bar{\phi}_{\mathrm{L}})$, both measured in the solar barycentric frame, as binary parameters.
Therefore we need to express the waveforms (especially $\hat{\bm{L}}\cdot\hat{\bm{N}}$ and the beam-pattern functions $F_{\alpha}^{+}$ and $F_{\alpha}^{\times}$ which appear in Eqs.~(\ref{Apol})-(\ref{phipol})) in terms of $\bar{\theta}_{\mathrm{S}},\bar{\phi}_{\mathrm{S}},\bar{\theta}_{\mathrm{L}}$ and $\bar{\phi}_{\mathrm{L}}$.
$\theta_{\mathrm{S}}(t)$ and $\phi_{\mathrm{S}}(t)$ are expressed as

\begin{widetext}
\begin{eqnarray}
\cos \theta_{\mathrm{S}}(t)&=&\frac{1}{2}\cos \bar{\theta_{\mathrm{S}}}
                                          -\frac{\sqrt{3}}{2}\sin \bar{\theta_{\mathrm{S}}} \cos[\bar{\phi}(t)-\bar{\phi_{\mathrm{S}}}],  \\
\phi_{\mathrm{S}}(t)&=&\alpha_1+\frac{\pi}{12}+\tan^{-1}
                              \left( \frac{\sqrt{3}\cos{\bar{\theta_{\mathrm{S}}}}
                              +\sin \bar{\theta_{\mathrm{S}}}\cos [\bar{\phi}(t)-\bar{\phi_{\mathrm{S}}}]}
                              {2\sin \bar{\theta_{\mathrm{S}}}\sin [\bar{\phi}(t)-\bar{\phi_{\mathrm{S}}}]} \right).
\end{eqnarray}
%\end{widetext}
For the polarisation angle $\psi_{\mathrm{S}}$ (see Eq.~(\ref{tanpsi})), first $\hat{\bm{z}}\cdot\hat{\bm{N}}=\cos\theta_{\mathrm{S}}$.
Next when we neglect the spin precessional effects, $\hat{\bm{L}}$ is constant and $\hat{\bm{L}}\cdot\hat{\bm{z}}$, $\hat{\bm{L}}\cdot\hat{\bm{N}}$, and $\hat{\bm{N}}\cdot(\hat{\bm{L}}\times\hat{\bm{z}})$ are given as

%\begin{widetext}
\begin{eqnarray}
\hat{\bm{L}}\cdot\hat{\bm{z}}&=&\frac{1}{2}\cos\bar{\theta}_{\mathrm{L}}
                   -\frac{\sqrt{3}}{2}\sin\bar{\theta}_{\mathrm{L}} \cos[\bar{\phi}(t)-\bar{\phi}_{\mathrm{L}}], \label{lz} \\
\hat{\bm{L}}\cdot\hat{\bm{N}}&=&\cos\bar{\theta}_{\mathrm{L}}\cos\bar{\theta}_{\mathrm{S}}
                  +\sin\bar{\theta}_{\mathrm{L}}\sin\bar{\theta}_{\mathrm{S}}\cos(\bar{\phi}_{\mathrm{L}}-\bar{\phi}_{\mathrm{S}}), \label{ln} \\
\hat{\bm{N}}\cdot(\hat{\bm{L}}\times\hat{\bm{z}})&=&\frac{1}{2}\sin\bar{\theta}_{\mathrm{L}}\sin\bar{\theta}_{\mathrm{S}}
                                             \sin(\bar{\phi}_{\mathrm{L}}-\bar{\phi}_{\mathrm{S}}) \notag \\
                                             &&-\frac{\sqrt{3}}{2}\cos\bar{\phi}(t)(
                                             \cos\bar{\theta}_{\mathrm{L}}\sin\bar{\theta}_{\mathrm{S}}\sin \bar{\phi}_{\mathrm{S}} 
                                             -\cos\bar{\theta}_{\mathrm{S}}\sin\bar{\theta}_{\mathrm{L}}\sin \bar{\phi}_{\mathrm{L}} ) \notag \\
                                             &&-\frac{\sqrt{3}}{2}\sin\bar{\phi}(t)(
                                             \cos\bar{\theta}_{\mathrm{S}}\sin\bar{\theta}_{\mathrm{L}}\cos \bar{\phi}_{\mathrm{L}} 
                                             -\cos\bar{\theta}_{\mathrm{L}}\sin\bar{\theta}_{\mathrm{S}}\cos \bar{\phi}_{\mathrm{S}} ). \label{nlz}
\end{eqnarray}
\end{widetext}

\section{THE ANALYTIC FORMULAE FOR THE ORBITAL ANGULAR MOMENTUM $\bm{L}$ UNDER SIMPLE PRECESSION}
\label{app-prec}

The precession equations for circular orbit binaries are~\cite{apostolatos} 

\begin{widetext}
\begin{eqnarray}
\dot{\bm{L}}&=&\frac{1}{r^3}\left[ \frac{4m_1+3m_2}{2m_1}\bm{S}_1+\frac{4m_2+3m_1}{2m_2}\bm{S}_2 \right] 
                           \times\bm{L} \notag \\
                          && -\frac{3}{2}\frac{1}{r^3}[(\bm{S}_2\cdot\hat{\bm{L}})\bm{S}_1
                            +(\bm{S}_1\cdot\hat{\bm{L}})\bm{S}_2]\times\hat{\bm{L}}-\frac{32}{5}\frac{\mu^2}{r}
                            \left( \frac{M^{5/2}}{r} \right) \hat{\bm{L}},  \label{Lprec} \\
\dot{\bm{S}}_1&=&\frac{1}{r^3}\left[ \frac{4m_1+3m_2}{2m_1}(\mu \sqrt{Mr}\hat{\bm{L}}) \right]\times \bm{S}_1
                              +\frac{1}{r^3}\left[ \frac{1}{2}\bm{S}_2-\frac{3}{2}(\bm{S}_2\cdot\hat{\bm{L}})\hat{\bm{L}} \right]
                              \times\bm{S}_1, \label{S1prec} \\ 
\dot{\bm{S}}_2&=&\frac{1}{r^3}\left[ \frac{4m_2+3m_1}{2m_2}(\mu \sqrt{Mr}\hat{\bm{L}}) \right]\times \bm{S}_2
                              +\frac{1}{r^3}\left[ \frac{1}{2}\bm{S}_1-\frac{3}{2}(\bm{S}_1\cdot\hat{\bm{L}})\hat{\bm{L}} \right]
                              \times\bm{S}_2. \label{S2prec} 
\end{eqnarray}
\end{widetext}
The first term of each equation represents the spin-orbit interactions (1.5PN) and the second term represents the spin-spin interactions (2PN).
The last term of Eq.~(\ref{Lprec}) is the angular momentum loss due to the radiation reaction.
This changes the total angular momentum $\bm{J}\equiv \bm{L}+\bm{S}_1+\bm{S}_2$ as

\begin{equation}
\dot{\bm{J}}=-\frac{32}{5}\frac{\mu^2}{r}\left( \frac{M^{5/2}}{r} \right) \hat{\bm{L}}.
\end{equation}

In this paper, we assume that one of the spins of the binary constituents is negligible (i.e. $\bm{S}_1\sim 0$).
Then, there do not exist spin-spin interactions.
We also assume that the orbital angular momentum $\bm{L}$ is neither parallel nor antiparallel to the total spin angular momentum $\bm{S}$($=\bm{S}_1+\bm{S}_2$).
Then, the precession equations are simplified and $\hat{\bm{L}}$ is obtained analytically up to some approximate orders.
This is the so-called \textit{simple precession approximation}~\cite{apostolatos}.
This also holds when the masses of the binary constituents are equal ($m_1\sim m_2$) and spin-spin interactions are negligible, instead of $\bm{S}_1\sim 0$.
Under this simple precession approximation, the precession equations become Eqs.~(\ref{40a})-(\ref{40d}).

Next, we define a quantity $\gamma (t)$ as 

\begin{equation}
\gamma(t) \equiv \frac{S}{L(t)}.
\end{equation}
Then, the magnitude and the direction of the total angular momentum $\bm{J}$ can be expressed in terms of  $\kappa$, $\gamma(t)$, $L(t)$, $\hat{\bm{L}}$ and $\hat{\bm{S}}$,

\begin{eqnarray}
J&=&L\sqrt{1+2\kappa \gamma+\gamma^2}, \label{48a} \\
\hat{\bm{J}}&=&\frac{\hat{\bm{L}}+\gamma \hat{\bm{S}}}{\sqrt{1+2\kappa \gamma+\gamma^2}}. \label{48b}
\end{eqnarray}
From Eqs.~(\ref{40c}),~(\ref{40d}) and~(\ref{48b}), the precession equation of $\hat{\bm{J}}$ can be derived as

\begin{equation}
\dot{\hat{\bm{J}}}=\frac{\dot{\gamma}[\hat{\bm{S}}(1+\kappa\gamma)-\hat{\bm{L}}(\kappa+\gamma)]}
                                     {(1+2\kappa \gamma+\gamma^2)^{3/2}}. \label{Jdothat}
\end{equation}

\begin{figure}[t]
 \begin{center}
  \includegraphics[scale=.8,clip]{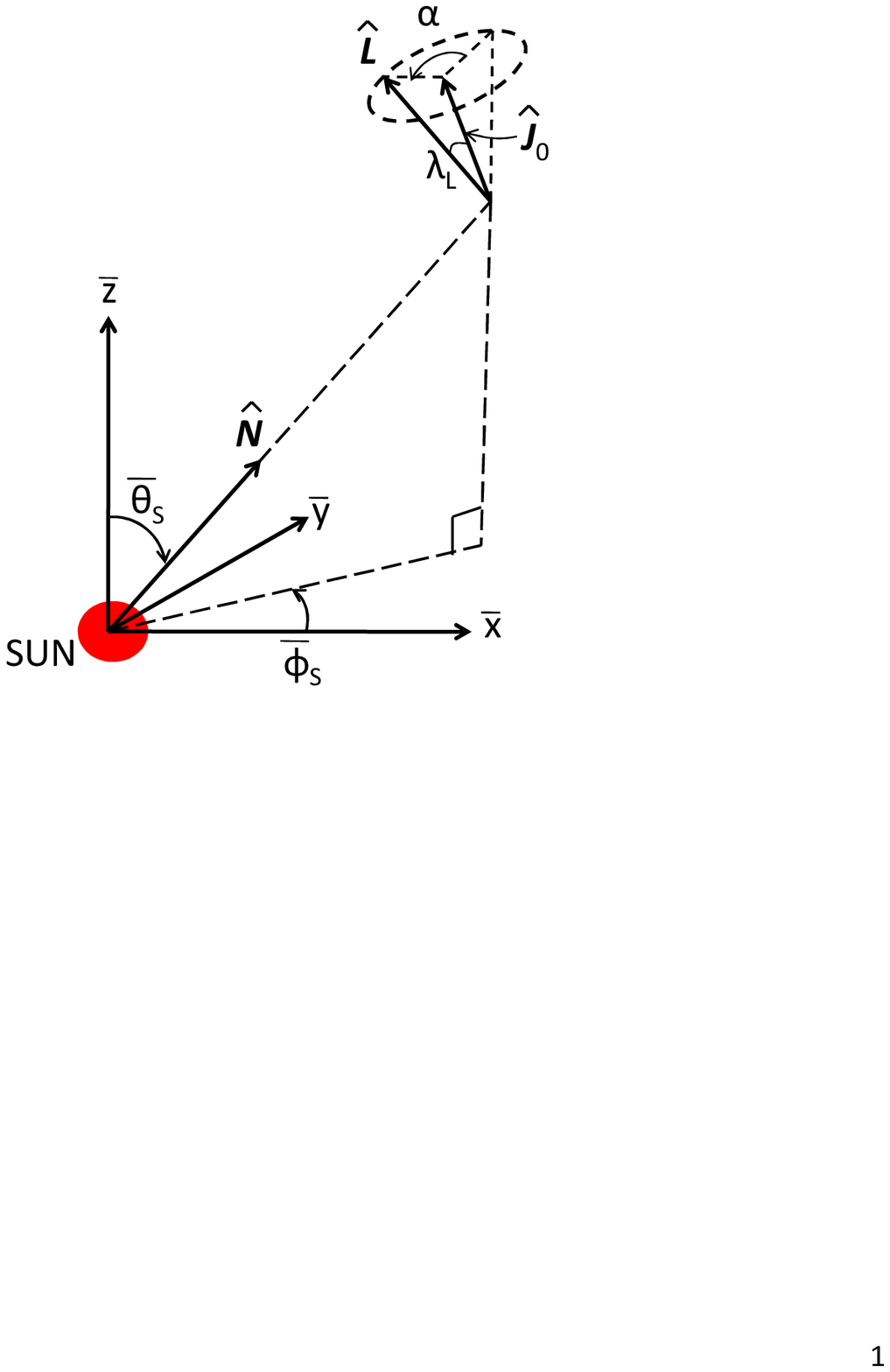} 
   \end{center}
 \caption{The unit orbital angular momentum vector $\hat{\bm{L}}$ precesses around a fixed vector $\hat{\bm{J}}_0$. The opening angle of the precession cone is given by $\lambda_{\mathrm{L}}$ and the precession angle is denoted as $\alpha$.}
\label{fig2}
\end{figure}

From Eqs.~(\ref{40c}) and~(\ref{40d}), it can be seen that the vectors $\hat{\bm{L}}$ and $\hat{\bm{S}}$ precess around $\hat{\bm{J}}$ with the angular velocity $\Omega_p$ given as Eq.~(\ref{omega-precess}).
In general, the precessing time scale $\Omega_p^{-1}$ is shorter than the inspiral time scale $L/\lvert \dot{L}\rvert$.
Therefore from $\dot{\bm{J}}=\dot{L}\hat{\bm{L}}$, $\bm{J}$ changes in magnitude but the direction is almost constant. Actually, if $J$ is much smaller than $L$ (as this can happen when $\bm{L}$ and $\bm{S}$ are antialigned with almost the same magnitudes), $\hat{\bm{J}}$ can change significantly in one precessional period.
Therefore, we introduce the following small parameter, 

\begin{equation}
\varepsilon\equiv\frac{L}{J}\frac{\lvert \dot{L} \rvert/L}{\Omega_p}=\frac{\lvert \dot{\bm{J}} \rvert/J}{\Omega_p}.
\end{equation}
Then, $\bm{J}$ precesses around the fixed direction $\hat{\bm{J}}_0$ with

\begin{equation}
\hat{\bm{J}}_0=\hat{\bm{J}}-\varepsilon\hat{\bm{J}}\times\hat{\bm{L}}
\end{equation}
up to $O(\varepsilon^2)$.
To the same order, the precession equation for $\hat{\bm{L}}$ becomes

\begin{equation}
\dot{\hat{\bm{L}}}=\Omega_p\hat{\bm{J}}_0\times\hat{\bm{L}}
                                 +\varepsilon\Omega_p(\hat{\bm{J}}_0\times\hat{\bm{L}})\times\hat{\bm{L}}. \label{57}
\end{equation}
The solution of this equation can be obtained geometrically~\cite{apostolatos}.
We take the barycentric Cartesian frame $(\bar{x},\bar{y},\bar{z})$ which is tied to the ecliptic and centred in the solar system barycentre.
Let the barycentre of the binary point in the $((\bar{\theta}_{\mathrm{S}},\bar{\phi}_{\mathrm{S}}))$ direction and let $\hat{\bm{J}}_0$ point in the $(\bar{\theta}_{\mathrm{J}},\bar{\phi}_{\mathrm{J}})$ direction.
We denote $\lambda_{\mathrm{L}}$ as the opening angle of the cone on which $\hat{\bm{L}}$ precesses (i.e. the angle between $\hat{\bm{L}}$ and $\hat{\bm{J}}_0$; see Fig.~\ref{fig2}).
This can be regarded as the angle between $\hat{\bm{L}}$ and $\hat{\bm{J}}$ apart from the errors of order $\varepsilon^2$ and is given by

\begin{eqnarray}
\cos\lambda_{\mathrm{L}}&=&\hat{\bm{L}}\cdot\hat{\bm{J}}=\frac{1+\kappa \gamma}{\sqrt{1+2\kappa \gamma+\gamma^2}}, \\
\sin\lambda_{\mathrm{L}}&=&\frac{\lvert \dot{\hat{\bm{L}}} \rvert}{\Omega_p}
                                      =\frac{\gamma\sqrt{1-\kappa^2}}{\sqrt{1+2\kappa \gamma+\gamma^2}}.
\end{eqnarray}
Then, $\hat{\bm{L}}$ can be expressed as

\begin{widetext}
\begin{equation}
\hat{\bm{L}}=\hat{\bm{J}}_0\cos\lambda_{\mathrm{L}}
                         +\frac{\hat{\bar{\bm{z}}}-\hat{\bm{J}}_0\cos\bar{\theta}_{\mathrm{J}}}{\sin\bar{\theta}_{\mathrm{J}}}
                           \sin\lambda_{\mathrm{L}}\cos\alpha
                         +(\hat{\bm{J}}_0\times\hat{\bar{\bm{z}}})\frac{\sin\lambda_{\mathrm{L}}\sin\alpha}{\sin\bar{\theta}_{\mathrm{J}}}, \label{L-prec}
\end{equation}
%\end{widetext}
where $\alpha$ is the \textit{precession angle} defined as the solution of

\begin{equation}
\frac{d\alpha}{dt}\equiv\Omega_p. \label{alpha-omega}
\end{equation} 
We assume that $\alpha=0$ is when $\hat{\bm{L}}\cdot\hat{\bar{\bm{z}}}$ is maximum (see Fig.~\ref{fig2}).
By solving the above equation, we get

%\begin{widetext}
\begin{equation}
\begin{split}
\alpha=&\alpha_c-\frac{5}{96}\frac{1}{\mu^3M^3}\left( 1+\frac{3}{4}\frac{m_1}{m_2} \right) \\
           &\times\left[ 2(\mathcal{G}L)^3-3\kappa S(L+\kappa S)\mathcal{G}L
             -3\kappa S^3(1-\kappa^2)\mathrm{arcsinh}\left( \frac{L+\kappa S}{S\sqrt{1-\kappa^2}}\right) \right],
\end{split}
\end{equation}
%\end{widetext}
where $\alpha_c$ is a quantity which characterises $\alpha$ at $t=t_c$ and $\mathcal{G}$ is defined as

\begin{equation}
\mathcal{G}\equiv \sqrt{1+2\kappa\gamma+\gamma^2}.
\end{equation}

From Eq.~(\ref{L-prec}), the quantities $(\hat{\bm{L}}\cdot\hat{\bm{N}})$,$(\hat{\bm{L}}\cdot\hat{\bm{z}})$ and $[\hat{\bm{N}}\cdot(\hat{\bm{L}}\times\hat{\bm{z}})]$, which are needed to compute the polarisation angle $\psi_{\mathrm{S}}(t)$ in the beam-pattern coefficients $F_{\alpha}^+$ and $F_{\alpha}^{\times}$, are expressed as~\cite{vecchio}

%\begin{widetext}
\begin{eqnarray}
\hat{\bm{L}}\cdot\hat{\bm{z}}&=&(\hat{\bm{J}}_0\cdot\hat{\bm{z}})\cos\lambda_{\mathrm{L}}
                     +\frac{1-2(\hat{\bm{J}}_0\cdot\hat{\bm{z}})\cos\bar{\theta}_{\mathrm{J}}}
                     {2\sin\bar{\theta}_{\mathrm{J}}}\sin\lambda_{\mathrm{L}}\cos\alpha \notag \\
                     & &+\frac{(\hat{\bm{J}}_0\times\hat{\bar{\bm{z}}})\cdot\hat{\bm{z}}}{\sin\bar{\theta}_{\mathrm{J}}}
                     \sin\lambda_{\mathrm{L}}\sin\alpha, \label{55} \\
\hat{\bm{L}}\cdot\hat{\bm{N}}&=&(\hat{\bm{J}}_0\cdot\hat{\bm{N}})\cos\lambda_{\mathrm{L}}
                     +\frac{\cos\bar{\theta}_{\mathrm{S}}-(\hat{\bm{J}}_0\cdot\hat{\bm{N}})\cos\bar{\theta}_{\mathrm{J}}}
                     {\sin\bar{\theta}_{\mathrm{J}}}\sin\lambda_{\mathrm{L}}\cos\alpha \notag \\
                     & &+\frac{(\hat{\bm{J}}_0\times\hat{\bar{\bm{z}}})\cdot\hat{\bm{N}}}{\sin\bar{\theta}_{\mathrm{J}}}
                     \sin\lambda_{\mathrm{L}}\sin\alpha, \label{54} \\
\hat{\bm{N}}\cdot(\hat{\bm{L}}\times\hat{\bm{z}})&=
                     &\hat{\bm{N}}\cdot (\hat{\bm{J}}_0\times\hat{\bm{z}})\cos\lambda_{\mathrm{L}}
                     +\frac{\hat{\bm{N}}\cdot(\hat{\bar{\bm{z}}}\times\hat{\bm{z}})
                     -\hat{\bm{N}}\cdot(\hat{\bm{J}}_0\times\hat{\bm{z}})\cos\bar{\theta}_{\mathrm{J}}}
                     {\sin\bar{\theta}_{\mathrm{J}}}\sin\lambda_{\mathrm{L}}\cos\alpha \notag \\
                     & &+\frac{\hat{\bm{N}}\cdot (\hat{\bm{J}}_0\times\hat{\bar{\bm{z}}})\times\hat{\bm{z}}}
                     {\sin\bar{\theta}_{\mathrm{J}}}
                     \sin\lambda_{\mathrm{L}}\sin\alpha, \label{56}
\end{eqnarray}
%\end{widetext}
where

%\begin{widetext}
\begin{eqnarray}
\hat{\bm{J}}_0\cdot\hat{\bm{z}}&=&\frac{1}{2}\cos\bar{\theta_{\mathrm{J}}}
                   -\frac{\sqrt{3}}{2}\sin\bar{\theta_{\mathrm{J}}} \cos[\bar{\phi}(t)-\bar{\phi_{\mathrm{J}}}], \\
\hat{\bm{J}}_0\cdot\hat{\bm{N}}&=&\cos\bar{\theta_{\mathrm{J}}}\cos\bar{\theta_{\mathrm{S}}}
                    +\sin\bar{\theta_{\mathrm{J}}}\sin\bar{\theta_{\mathrm{S}}}\cos(\bar{\phi_{\mathrm{J}}}-\bar{\phi_{\mathrm{S}}}), \label{Jn} \\
\hat{\bm{N}}\cdot (\hat{\bm{J}}_0\times\hat{\bm{z}})&=&\frac{1}{2}\sin\bar{\theta_{\mathrm{J}}}\sin\bar{\theta_{\mathrm{S}}}
                                             \sin(\bar{\phi_{\mathrm{J}}}-\bar{\phi_{\mathrm{S}}}) \notag \\
                                             &&-\frac{\sqrt{3}}{2}\cos\bar{\phi(t)}(
                                             \cos\bar{\theta_{\mathrm{J}}}\sin\bar{\theta_{\mathrm{S}}}\sin \bar{\phi_{\mathrm{S}}} 
                                             -\cos\bar{\theta_{\mathrm{S}}}\sin\bar{\theta_{\mathrm{J}}}\sin \bar{\phi_{\mathrm{J}}} ) \notag \\
                                             &&-\frac{\sqrt{3}}{2}\sin\bar{\phi(t)}(
                                             \cos\bar{\theta_{\mathrm{S}}}\sin\bar{\theta_{\mathrm{J}}}\cos \bar{\phi_{\mathrm{J}}} 
                                             -\cos\bar{\theta_{\mathrm{J}}}\sin\bar{\theta_{\mathrm{S}}}\cos \bar{\phi_{\mathrm{S}}} ), \label{nJz} \\
(\hat{\bm{J}}_0\times\hat{\bar{\bm{z}}})\cdot\hat{\bm{N}}&=&\sin\bar{\theta}_{\mathrm{S}}\sin\bar{\theta}_{\mathrm{J}}
                             \sin (\bar{\phi}_{\mathrm{J}}-\bar{\phi}_{\mathrm{S}}), \\
\hat{\bm{N}}\cdot(\hat{\bar{\bm{z}}}\times\hat{\bm{z}})&=&
                       \frac{\sqrt{3}}{2}\sin\bar{\theta}_{\mathrm{S}}\sin(\bar{\phi}(t)-\bar{\phi}_{\mathrm{S}}), \\
(\hat{\bm{J}}_0\times\hat{\bar{\bm{z}}})\cdot\hat{\bm{z}}&=&
                        \frac{\sqrt{3}}{2}\sin\bar{\theta}_{\mathrm{J}}\sin(\bar{\phi}(t)-\bar{\phi}_{\mathrm{J}}), \\
\hat{\bm{N}}\cdot (\hat{\bm{J}}_0\times\hat{\bar{\bm{z}}})\times\hat{\bm{z}}&=&
                  -\frac{1}{2}\sin\bar{\theta}_{\mathrm{J}}[ \sqrt{3}\cos\bar{\theta}_{\mathrm{S}}\cos\{ \bar{\phi}(t)-\bar{\phi}_{\mathrm{J}} \}                  +\sin\bar{\theta}_{\mathrm{S}}\cos(\bar{\phi}_{\mathrm{J}}-\bar{\phi}_{\mathrm{S}}). 
\end{eqnarray}
When spins are zero, $\sin\lambda_{\mathrm{L}}=0,\bar{\theta}_{\mathrm{J}}=\bar{\theta}_{\mathrm{L}},
\bar{\phi}_{\mathrm{J}}=\bar{\phi}_{\mathrm{L}}$.
Then, Eqs.~(\ref{55}),~(\ref{54}) and~(\ref{56}) each reduces to Eqs.~(\ref{lz}),~(\ref{ln}) and~(\ref{nlz}), respectively.

$\hat{\bm{L}}\cdot\hat{\bm{u}}$ and $\hat{\bm{N}}\cdot(\hat{\bm{L}}\times\hat{\bm{u}})$ in Eq.~(\ref{thomas-new}) are given as follows:

\begin{eqnarray}
\hat{\bm{L}}\cdot\hat{\bm{u}}&=&-\hat{\bm{N}}\cdot(\hat{\bm{L}}\times\hat{\bm{J}}_0) \\ \notag
                                                       &=&\hat{\bm{N}}\cdot(\hat{\bm{J}}_0\times\hat{\bm{z}})
                              \frac{\sin\lambda_{\mathrm{L}}}{\sin\theta_{\mathrm{J}}}\cos\alpha
                              -(\cos\theta_{\mathrm{S}}-\cos\theta_{\mathrm{J}}(\hat{\bm{J}}_0\cdot\hat{\bm{N}}))
                              \frac{\sin\lambda_{L}}{\sin\theta_{\mathrm{J}}}\sin\alpha, \\
\hat{\bm{N}}\cdot(\hat{\bm{L}}\times\hat{\bm{u}}) &= &\cos \lambda_{\mathrm{L}}
                               -(\hat{\bm{L}}\cdot\hat{\bm{N}})\cdot(\hat{\bm{J}}_0\cdot\hat{\bm{N}}),
\end{eqnarray}
where $\hat{\bm{N}}\cdot(\hat{\bm{J}}_0\times\hat{\bm{z}})$ and
$\hat{\bm{J}}_0\cdot\hat{\bm{N}}$ are given as Eqs.~(\ref{nJz}) and~(\ref{Jn}), respectively.
\end{widetext}

\section{THE INVERSION OF THE FISHER MATRIX}
\label{app-inv}

\quad If the ratio (we denote this by $R$) of the smallest eigenvalue to the largest one in the Fisher matrix $\mathbf{\Gamma}$ approaches the machine's floating-point precision, the inverse of $\mathbf{\Gamma}$ will not be performed correctly.
This problem can be avoided by the following technique.

First, we define 

\begin{equation}
\mathbf{T}\equiv \mathrm{diag}\left(\frac{1}{\sqrt{\Gamma_{ii}}}\right).
\end{equation}
Next, we obtain the normalised Fisher matrix $\mathbf{\Gamma}'$ as follows:

\begin{equation}
\mathbf{\Gamma}'\equiv \mathbf{T}\mathbf{\Gamma}\mathbf{T}^{\mathrm{T}}=\mathbf{T}\mathbf{\Gamma}\mathbf{T}.
\end{equation} 
Then, all the diagonal components of $\mathbf{\Gamma}'$ equal to 1.
After that, we take the inverse of $\mathbf{\Gamma}'$.
From $\mathbf{\Gamma}'^{-1}=\mathbf{T}^{-1}\mathbf{\Gamma}^{-1}\mathbf{T}^{-1}$, we obtain the inverse of our original Fisher matrix $\mathbf{\Gamma}$ by multiplying $\mathbf{T}$ from both sides of $\mathbf{\Gamma}'^{-1}$,

\begin{equation}
\mathbf{\Gamma}^{-1}=\mathbf{T}\mathbf{\Gamma}'^{-1}\mathbf{T}.
\end{equation}
Even if the ratio $R$ is smaller than the machine's floating-point precision, the ratio for the normalised Fisher matrix $\mathbf{\Gamma}'$ can be larger than the floating-point precision so that the inversion can be performed correctly even in the double precision computation.

To check that our inversion is correctly performed, we followed a similar procedure described in the Appendix of Berti \textit{et al.}~\cite{berti}.
We simply multiply the inversed Fisher matrix by the original one so that we obtain a numerical ``identity matrix" $I_{ij}^{\mathrm{num}}$.
 Then, we define the following small quantity
\begin{equation}
\varepsilon_{\mathrm{inv}}\equiv \max_{i,j}\Big|\sqrt{I_{ij}^{\mathrm{num}\cdot}I_{ji}^{\mathrm{num}}}-\delta_{ij}\Big|.
\end{equation} 
With our inversion, we found that $\varepsilon_{\mathrm{inv}}$ easily satisfies the criteria, 
$\varepsilon_{\mathrm{inv}}<10^{-3}$ (for NS/BH binaries) and $\varepsilon_{\mathrm{inv}}<10^{-4}$ (for BH/BH binaries),  which are used in Ref.~\cite{berti}.

\section{\label{app1}The evolution of ECCENTRICITY}

\quad In this section, we explain how to calculate eccentricity $e$ at a given frequency $f$, with some initial eccentricity $e=e_i$ at a frequency $f=f_i$.
According to Peters~\cite{peters}, the evolution equations for the semimajor axis $a$ and the eccentricity $e$ of the orbit are given by

\begin{eqnarray}
\frac{da}{dt}&=&-\frac{64}{5}\frac{m_1m_2M}{a^3(1-e^2)^{7/2}}\left( 1+\frac{73}{24}e^2+\frac{37}{96}e^4 \right), \label{dadt} \\ 
\frac{de}{dt}&=&-\frac{304}{15}e\frac{m_1m_2M}{a^4(1-e^2)^{5/2}}\left( 1+\frac{121}{304}e^2 \right). \label{dedt}
\end{eqnarray}
Dividing Eq.~(\ref{dadt}) by Eq.~(\ref{dedt}), we get

\begin{equation}
\frac{da}{a}=\frac{19}{12}\frac{1+\frac{73}{24}e^2+\frac{37}{96}e^4}{e(1-e^2)( 1+\frac{121}{304}e^2 )}de.
\end{equation} 
Then, we integrate this equation to give

\begin{equation}
\frac{a}{a_i}=\frac{1-e_i^2}{1-e^2}\left( \frac{e}{e_i} \right)^{12/19}
                   \left[ \frac{ 1+\frac{121}{304}e^2 }{ 1+\frac{121}{304}e_i^2 } \right]^{\frac{870}{2299}}, \label{a(e)}
\end{equation}
where the subscript $i$ denotes the value at the initial time.
When $e/e_i \ll 1$, Eq.~(\ref{a(e)}) becomes

\begin{equation}
e\sim\left( \frac{a( 1+\frac{121}{304}e_i^2)}{a_i(1-e_i^2)}  \right)^{19/12}e_i. \label{eccentricity}
\end{equation}

Next, we calculate the time to coalescence $T_c(a_i,e_i)$.
This can be derived from substituting Eq.~(\ref{a(e)}) into Eq.~(\ref{dedt}), taking the reciprocal and performing the integration with $e$ from 0 to $e_i$,

\begin{widetext}
\begin{equation}
%\begin{split}
T_c(a_i,e_i)\equiv \int^{0}_{e_i} \frac{dt}{de}de
                =\frac{15}{304}\frac{a_i^4}{m_1m_2M}e_i^{-48/19}\frac{(1-e_i^2)^4}{(1+\frac{121}{304}e_i^2)^{\frac{3480}{2299}}}
                    \int^{e_i}_0 e^{29/19}\frac{(1+\frac{121}{304}e^2)^{\frac{1181}{2299}}}{(1-e^2)^{3/2}}de. \label{tca0e0}
%\end{split}
\end{equation}
%\end{widetext}
In the limit of $e_i \rightarrow 0$, the equation above reduces to

\begin{equation}
%\begin{split}
T_c(a_i)=\frac{15}{304}\frac{a_i^4}{m_1m_2M}e_i^{-48/19}\int^{e_i}_0 e^{29/19}de 
          =\frac{5}{256}\frac{a_i^4}{m_1m_2M}. \label{tca0}
%\end{split}
\end{equation}
Then, Eq.~(\ref{tca0e0}) can be expressed as 

%\begin{widetext}
\begin{equation}
T_c(a_i,e_i)=\frac{48}{19}T_c(a_i)e_i^{-48/19}\frac{(1-e_i^2)^4}{(1+\frac{121}{304}e_i^2)^{\frac{3480}{2299}}}
                    \int^{e_i}_0 e^{29/19}\frac{(1+\frac{121}{304}e^2)^{\frac{1181}{2299}}}{(1-e^2)^{3/2}}de.
\end{equation} 
\end{widetext}
When $e_i\sim 1$, this equation can be written as

\begin{equation}
T_c(a_i,e_i) \approx \frac{768}{425}T_c(a_i)(1-e_i^2)^{7/2}. \label{tca0e02}
\end{equation}

From these equations, we can calculate the eccentricity $e$ at frequency $f$ for the binary whose initial eccentricity is $e_i\sim 1$ and coalesces in time $t(f)=\frac{5}{256}\mathcal{M} (\pi \mathcal{M} f)^{-8/3}$ (see Eq.~(\ref{tf})).  
From Eqs.~(\ref{tca0}) and~(\ref{tca0e02}), we eliminate $T_c(a_i)$ to get

\begin{equation}
a_i \simeq \left( \frac{256}{5}\frac{T_c(a_i,e_i)\mu M^2}{(1-e_i^2)^{7/2}} \right)^{1/4}, \label{ai}
\end{equation}
where $T_c(a_i,e_i)$ is given by $T_c(a_i,e_i)=t(f_i)$.
On the other hand, when $e\ll 1$, we can replace the semimajor axis $a(f)$ at frequency $f$ by the orbital separation $r(t(f))$ as~\cite{flanagan}

\begin{equation}
a(f)\approx r(t(f))=\left( \frac{256}{5}\mu M t(f) \right)^{1/4}. \label{a0}
\end{equation}
Then, we obtain our value $e$ by substituting Eqs.~(\ref{ai}) and (\ref{a0}) into Eq.~(\ref{eccentricity}).
For (1.4+10$^3$)$M_{\odot}$ NS/BH binary with initial eccentricity $e_i=0.998$ at $f_i=2\times 10^{-4}$Hz, the eccentricity $e$ at frequency $f=f_{\mathrm{1yr}}$ becomes $e=0.026$.

\section{THE ANALYTIC FORMS OF $\frac{\partial \tilde{h}}{\partial \theta^a}$ USED IN THE PATTERN-AVERAGED ESTIMATES}
\label{derivative}

\quad In Sec.~\ref{sec-num-ave}, we calculate the parameter determination errors with pattern-averaged estimates. There are 9 parameters in total: $\ln \mathcal{M},\ln \eta, t_c,\phi_c,D_L,\beta,\sigma,e_0$, and $\bar{\omega}$ or $\beta_g$.
The waveform is given by Eq.~(\ref{wave-noangle}), where the amplitude $\mathcal{A}$ and the phase $\Psi(f)$ are given by Eqs.~(\ref{amp-noangle}) and~(\ref{Psi-noangle}), respectively.
The derivative of this waveform with respect to each parameter is taken analytically as~\cite{berti}

%\begin{widetext}
\begin{eqnarray}
\frac{\partial\tilde{h}}{\partial\ln\mathcal{M}}&=&-\frac{5i}{128}(\pi\mathcal{M}f)^{-5/3}(k_4x^{-1}+1+d_4 x^{-19/6} \notag \\
                & &+a_4x+b_4x^{3/2}+c_4x^2)\tilde{h}, \\
\frac{\partial\tilde{h}}{\partial\ln\mathcal{\eta}}&=&-\frac{i}{96}(\pi\mathcal{M}f)^{-5/3}(k_5x^{-1}+1+d_5 x^{-19/6} \notag \\
                & &+a_5x+b_5x^{3/2}+c_5x^2)\tilde{h}, \\
\frac{\partial\tilde{h}}{\partial\ln D_L}&=&-\tilde{h}, \\
\frac{\partial\tilde{h}}{\partial t_c}&=&i2\pi f \tilde{h}, \\
\frac{\partial\tilde{h}}{\partial \phi_c}&=&-i\tilde{h}, \\
%\end{eqnarray}
%\begin{eqnarray}
\frac{\partial\tilde{h}}{\partial \beta}&=&\frac{3i}{32}(\pi \mathcal{M}f)^{-5/3}x^{3/2} \tilde{h}, \\
\frac{\partial\tilde{h}}{\partial \sigma}&=&-\frac{15i}{64}(\pi \mathcal{M}f)^{-5/3}x^{2} \tilde{h}, \\
\frac{\partial\tilde{h}}{\partial I_e}&=&-\frac{7065i}{187136}(\pi \mathcal{M}f)^{-5/3}x^{-19/6} \tilde{h}, \\
\frac{\partial\tilde{h}}{\partial \bar{\omega}}&=&-\frac{5i}{3584}\mathcal{S}^2(\pi \mathcal{M}f)^{-5/3}x^{-1} \tilde{h}, \\
\frac{\partial\tilde{h}}{\partial \beta_g}&= &-i(\pi\mathcal{M}f)^{-1}\tilde{h},
\end{eqnarray}
where

\begin{eqnarray}
a_4&=&\frac{4}{3}\left( \frac{743}{336}+\frac{11}{4}\eta \right )-\frac{128}{5}\beta_g \eta^{2/5}, \\
b_4&=&\frac{8}{5}(\beta-4\pi), \\
c_4&=&2\left( \frac{3058673}{1016064}+\frac{5429}{1008}\eta+\frac{617}{144}\eta^2-\sigma \right), \\
d_4&=&-\frac{2355}{1462}I_e, \\
k_4&=&-\frac{\mathcal{S}^2}{12}\bar{\omega}, \\
a_5&=&\frac{743}{168}-\frac{33}{4}\eta, \\
b_5&=&\frac{27}{5}(\beta-4\pi), \\
c_5&=&18\left( \frac{3058673}{1016064}-\frac{5429}{4032}\eta-\frac{617}{96}\eta^2-\sigma \right), \\
d_5&=&\frac{26847}{5848}I_e, 
\end{eqnarray}

\begin{eqnarray}
k_5&=&\frac{3\mathcal{S}^2}{56}\bar{\omega}.
\end{eqnarray}
%\end{widetext}

%\newpage %Just because of unusual number of tables stacked at end


\begin{thebibliography}{99}
\bibitem{comment}
Just after we have completed our analysis, there appeared a paper by Stavridis and Will~\cite{stavridis}, in which they estimated the possible bound on graviton mass from SMBH/BH inspiral gravitational waves using LISA.
Following Lang and Hughes~\cite{lang}, they randomly distributed the spins of both binary constituents and solved the precession equations numerically.
They took spin-spin coupling $\sigma$ into account but they did not include eccentricity into parameters.
When neglecting the precessional spin-spin coupling effect, they obtained the probability distribution of $\lambda_g$ which peaks around $\lambda_g=4\times10^{21}$cm a BH/BH binary of $(10^6+10^6)M_{\odot}$ at 3Gpc.
To compare our calculation with their result, we performed another Monte Carlo simulation of $(10^6+1.1\times 10^6)M_{\odot}$ binaries. 
We set the spin of one of the black holes to 0 and we randomly distribute the spin of the other black hole from $\chi=0$ to 2.
($\chi$ is the dimensionless spin parameter defined in Sec.~\ref{sec-wave}.
Although $\chi >1$ is unphysical, we extended it up to $\chi =2$ just to compare our result with the one in Ref.~\cite{stavridis}.)
We included the spin-spin coupling $\sigma$ into binary parameters but did not take eccentricity into account.
We obtained the probability distribution which has a peak around $\lambda_g =3.7\times 10^{21}$cm.
%This result is consistent with the one in Ref.~\cite{stavridis}.   
Although we cannot directly compare these two results, it seems that our result is consistent with the one obtained by Stavridis and Will~\cite{stavridis}.
\end{thebibliography}

\end{document}